\documentclass{emulateapj}
\usepackage{gensymb}
\usepackage{upgreek}
\usepackage{color}

\def\micron{$\upmu$m}

\begin{document}

\shortauthors{Esplin \& Luhman}
\shorttitle{Survey for New Members of Taurus}

\title{A Survey for New Members of Taurus from Stellar to Planetary Masses\altaffilmark{1}}

\author{
T. L. Esplin\altaffilmark{2,3} and
K. L. Luhman\altaffilmark{4,5}
}

\altaffiltext{1}{Based on observations made with the NASA Infrared Telescope
Facility, Gemini Observatory, UKIRT, MMT, CFHT, Pan-STARRS1, 2MASS, UKIDSS, 
UHS, SDSS, {\it Gaia}, {\it WISE}, and the {\it Spitzer Space Telescope}, 
which is operated by the Jet Propulsion Laboratory, 
California Institute of Technology under a contract with NASA.}
\altaffiltext{2}{Steward Observatory, University of Arizona, Tucson, AZ, 85719, 
USA; taranesplin@email.arizona.edu}
\altaffiltext{3}{Strittmatter Fellow}
\altaffiltext{4}{Department of Astronomy and Astrophysics, The Pennsylvania
State University, University Park, PA 16802, USA}
\altaffiltext{5}{Center for Exoplanets and Habitable Worlds,
The Pennsylvania State University, University Park, PA 16802, USA}

\begin{abstract}

We present a large sample of new members of the Taurus star-forming region
that extend from stellar to planetary masses.
To identify candidate members at substellar masses, we have used
color-magnitude diagrams and proper motions measured with several
wide-field optical and infrared (IR) surveys.
At stellar masses, we have considered the candidate members that
were found in a recent analysis of high-precision astrometry from the
{\it Gaia} mission. 
Using new and archival spectra, we have measured spectral
types and assessed membership for these 161 candidates, 79 of 
which are classified as new members.  
Our updated census of Taurus now contains 519 known members.
 According to {\it Gaia} data, this census should be nearly 
complete for spectral types earlier than M6--M7 at $A_J<1$.
For a large field encompassing $\sim72$\% of the known members, the census
should be complete for $K<15.7$ at $A_J<1.5$, which corresponds to 
$\sim5$--13~$M_{\rm Jup}$ for ages of 1--10~Myr based on theoretical
evolutionary models. Our survey has doubled the number of known members at 
$\geq$M9 and has uncovered the faintest known member in $M_K$, which should
have a mass of $\sim$3--10~$M_{\rm Jup}$ for ages of 1--10~Myr.
We have used mid-IR photometry from the {\it Spitzer Space Telescope}
and the {\it Wide-field Infrared Survey Explorer} to determine 
whether the new members exhibit excess emission that would indicate the presence
of circumstellar disks. The updated disk fraction for Taurus is
$\sim$0.7 at $\leq$M3.5 and $\sim$0.4 at $>$M3.5.

\end{abstract}

\keywords{}

\section{Introduction}

The Taurus cloud complex is one of the nearest star-forming regions
\citep[$d\sim140$~pc,][references therein]{gal18} and has a relatively
large stellar population \citep[$N\sim500$,][this work]{ken08},
making it well-suited for studies of star formation
that reach low stellar masses and have good statistics.
In addition, Taurus has an unusually low stellar density 
compared to other nearby molecular clouds, so it can help constrain
how the star formation process depends on environment.
However, a complete census of Taurus is challenging given that
its members are distributed across a large area of sky
($\sim100$~deg$^2$).

Surveys for members of Taurus have steadily improved in sensitivity
and areal coverage over the last 30 years 
\citep[][references therein]{kra17,luh17}.
We have recently sought to advance this work in \citet{esp17} and \citet{luh18}.
In the first survey, we searched for members down to planetary
masses ($<15$~$M_{\rm Jup}$) across a large fraction of Taurus using optical 
and infrared (IR) imaging from the {\it Spitzer Space Telescope} 
\citep{wer04}, the United Kingdom Infrared Telescope (UKIRT) Infrared Deep 
Sky Survey \citep[UKIDSS,][]{law07}, Pan-STARRS1 \citep[PS1,][]{kai02,kai10},
and the {\it Wide-field Infrared Survey Explorer} \citep[{\it WISE},][]{wri10}.
Meanwhile, \citet{luh18} used high-precision astrometry and optical photometry
from the second data release (DR2) of the {\it Gaia} mission 
\citep{per01,deb12,gaia16b,gaia18} to perform a thorough census
of stellar members with low-to-moderate extinctions across the entire cloud
complex.

We have continued the survey for low-mass brown dwarfs in Taurus
from \citet{esp18} by including new IR imaging from UKIRT and the 
Canada-France-Hawaii Telescope (CFHT). We also have obtained spectra of
most of the candidate stellar members that were identified with {\it Gaia}
by \citet{luh18}.
In this paper, we update the catalog of known members of Taurus from
\citet{luh18} (Section~\ref{sec:mem}), identify candidate members
using photometry, proper motions, and parallaxes
(Sections~\ref{sec:ident1} and \ref{sec:ident2}),
and spectroscopically classify the candidates (Section \ref{sec:spec}).
We assess the new members for evidence of circumstellar disks and
estimate the disk fraction as a function of stellar mass among the
known members (Section~\ref{sec:disk}). We conclude by using our new
census of Taurus to constrain the region's initial mass function (IMF),
particularly at the lowest masses (Section~\ref{sec:imf}).

\section{Catalog of Known Members of Taurus}
\label{sec:mem}

For our census of Taurus, we begin with the 438 objects adopted as members
by \citet{luh18}, which were vetted for contaminants using the proper motions
and parallaxes from {\it Gaia} DR2.
In that catalog, the components of a given multiple system appeared as
a single entry if they were unresolved by {\it Gaia} and
the imaging data utilized by \citet{esp17}. 
\citet{luh18} overlooked the fact that HK~Tau~B and V1195~Tau~B were
resolved from their primaries by {\it Gaia}. They are now given separate
entries in our census. We also adopt 2MASS J04282999+2358482 as a member,
which was discovered to be a young late-M object by \citet{giz99}.
It satisfies our photometric and astrometric criteria for membership
and is located near other known members. 
We exclude from our membership list one of the stars from from \citet{luh18},
2MASS 05023985+2459337, for reasons discussed in the Appendix.
When the 79 new members from our survey are included (Section~\ref{sec:spec}),
we arrive at a catalog of 519 known members of Taurus, which
are presented in Table~\ref{tab:mem}.  
That tabulation contains adopted spectral types,
astrometry and photometry from {\it Gaia} DR2 and the corresponding kinematic
populations from \citet{luh18}, proper motions measured in Section~\ref{sec:pm},
near-IR photometry from various sources,
mid-IR photometry from {\it Spitzer} and {\it WISE} and the resulting
disk classifications \citep[][Section~\ref{sec:disk}]{luh10,esp14,esp17},
and extinction estimates. For each star that appears in {\it Gaia} DR2, 
we also list the value of the re-normalized unit weight error
\citep[RUWE,][]{lin18}, which indicates the quality of the astrometric fit
(Section~\ref{sec:ident2}). \citet{luh18} compiled available radial velocity
measurements for known members of Taurus and calculated 
$UVW$ velocities from the combination of those radial velocities
and the {\it Gaia} proper motions and parallaxes. Two of the new members
from our survey,
{\it Gaia} 146708734143437568 and 152104381299305856, also have radial velocity
measurements (16.6$\pm$0.8~km~s$^{-1}$, 17.5$\pm$1.6~km~s$^{-1}$), both of 
which are from {\it Gaia} DR2.

A map of the spatial distribution of the members is shown in 
Figure~\ref{fig:spatial1}. Kinematic and photometric data for the members
within nine fields that cover subsections of Taurus are plotted in
Figures~\ref{fig:spatial2}-\ref{fig:spatiallast}, which contain diagrams from
\citet{luh18} that have been updated to include the new members from this work.
The boundaries for those fields are indicated in Figure~\ref{fig:spatial1}.

\section{Identification of Candidate Members at Substellar Masses}
\label{sec:ident1}

\subsection{Photometry and Astrometry}

\subsubsection{Data Utilized by \citet{esp17}}
\label{sec:esp17}

\begin{figure}[h]
	\centering
	\includegraphics[trim = 0mm 0mm 0mm 0mm, clip=true, scale=.7]{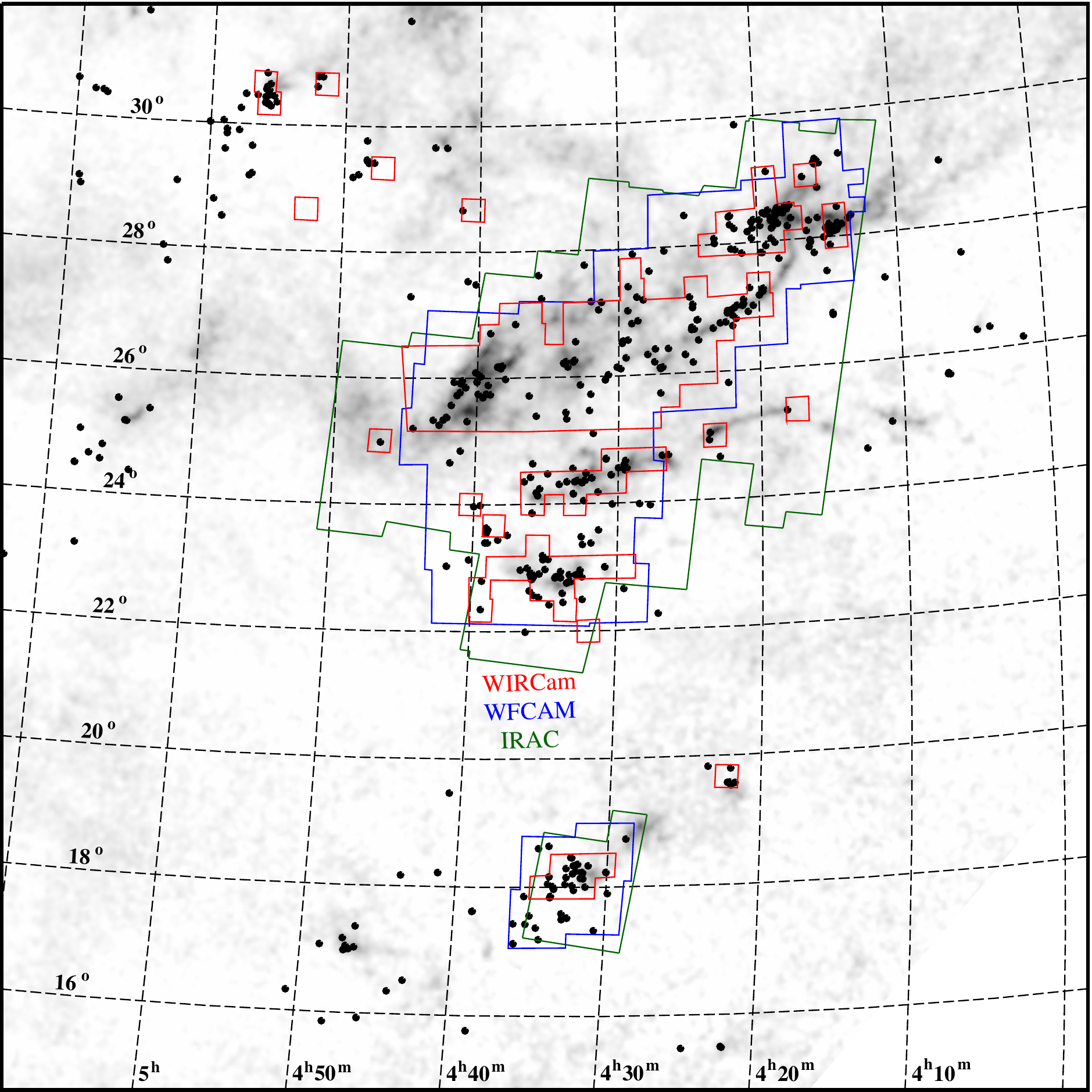}
\caption{
Fields in Taurus that have been imaged with IRAC, WIRCam, and in $JHK$ with
WFCAM (UKIDSS, UHS, new data).  The dark clouds are displayed with a map of
extinction \citep[gray scale;][]{dob05}.
}
\label{fig:fields}
\end{figure}

In \citet{esp17}, we identified candidate substellar members of Taurus
based on their proper motions and positions in color-magnitude diagrams 
(CMDs). We considered astrometry and photometry for objects within a field
encompassing all of the Taurus clouds 
($\alpha=4^{\rm h}$--$5^{\rm h}10^{\rm m}$,
$\delta=15\arcdeg$--31$\arcdeg$)
in several optical and IR bands:
$JHK_s$ from the Point Source Catalog of the Two Micron All Sky Survey 
\citep[2MASS,][]{cut03,skr06},
bands at 3.6, 4.5, 5.8, and 8.0~\micron\ ([3.6], [4.5], [5.8], [8.0])
from the Infrared Array Camera \citep[IRAC;][]{faz04} on
the {\it Spitzer Space Telescope} \citep{wer04},
$ZYJHK$ from data release 10 of UKIDSS,
$rizy_{P1}$ from the first data release of PS1 \citep{cha16,fle16},
$riz$ from data release 13 of the Sloan Digital Sky Survey 
\citep[SDSS,][]{yor00,fin04,alb17},
$G$ from the first data release of {\it Gaia} \citep{gaia16a,gaia16b},
and bands at 3.5, 4.6, 12, and 22~\micron\ ($W1$, $W2$, $W3$, $W4$)
from the AllWISE Source Catalog.
The extinction for each object was estimated using $J-H$ and $J-K_s$
colors and it was used to deredden the photometry in the CMDs with
reddening relations from \citet{ind05}, \citet{sch16a}, and \citet{xue16}.
We measured relative proper motions between 
2MASS and {\it Gaia} DR1, between 2MASS and PS1, and across several epochs
of IRAC imaging. Relative motions from UKIDSS were also employed.
2MASS, {\it WISE}, PS1 provided data for the entirety of our survey 
field while {\it Spitzer}, SDSS, and UKIDSS covered a subset of it
\citep{luh17,esp17}. The fields observed by IRAC are indicated in 
Figure~\ref{fig:fields}.

\subsubsection{UHS}

New $J$-band photometry has become available in Taurus through the first data
release of the UKIRT Hemisphere Survey \citep[UHS,][]{dye18}. 
Those data were taken with UKIRT's Wide Field Camera 
(WFCAM) \citep{cas01}, which also was used for UKIDSS. 
UHS provides $J$ photometry for a large portion of Taurus that was not observed
by UKIDSS in that band. We have adopted the $1\arcsec$ aperture photometry 
from UHS, which has a similar completeness limit as the data from UKIDSS
($J\sim18.5$).

\subsubsection{UKIRT}
\label{sec:ukirtphot}

UKIDSS has imaged most of Taurus in $K$ and roughly half of it in $ZYJ$.
Only a small portion of Taurus was observed in $H$.
When combined with $J$ and an optical band,
$H$ is particularly useful for distinguishing late-type members
from reddened background stars.
To improve the coverage of Taurus in $H$ and the other bands, 
we have obtained new images with WFCAM at UKIRT.
The observations were similar to those from UHS and UKIDSS, consisting of 
$3\times15$~s exposures in $Y$ and $4\times10$~s exposures in $JHK$ at
each position. The data were collected between September and December of 2017.
In Figure~\ref{fig:fields}, we show the fields that now have $JHK$ photometry
from WFCAM through UKIDSS, UHS, and our observations.

The initial data reduction steps (e.g., flat fielding, registration,
coaddition) were performed by the WFCAM pipeline \citep{irw04,ham08}.
We derived the flux calibration for the resulting images using photometry
from previous surveys (e.g., PS1, 2MASS).
The typical values of FWHM for point sources in the images were
$1\farcs1$ for $Y$ and $0\farcs8$ for $JHK$. 
We identified sources in the pipeline images and measured aperture photometry
for them using the routines {\tt starfind} and {\tt phot} in IRAF.
We estimated the completeness limits of the data based on the
magnitudes at which the logarithm of the number of stars as a function of
magnitudes deviates from a linear slope and begins to decline,
which were 18.5, 17.5, and 17.2 for $J$, $H$, and $K$, respectively.
Similar limits are exhibited by the UKIDSS data in Taurus. 
Our $Y$ data have a completeness limit near 18.5, which is $\sim0.5$~mag
brighter than the value for UKIDSS.

\subsubsection{CFHT}

Near-IR images of portions of Taurus are publicly available from the archive of 
CFHT. We have utilized the data taken in $J$, $H$, and a narrowband filter
at 1.45~\micron\ ($W$) with the Wide-field Infrared Camera (WIRCam) 
through programs 15BC11, 16AC13, 16BC17 (E. Artigau), 15BT11 (W.-P. Chen),
16AF16, 16BF22 (M. Bonnefoy), and 16AT04 (P. Chiang).
For most of the observations, individual exposure times were 10, 10,
and 65~s for $J$, $H$, and $W$, respectively, and the number of exposures per
field was 4--8. Point sources in the images typically exhibited
FWHM$\sim0\farcs6$--$0\farcs8$.
We began our analysis with the images from the CFHT archive that had been
pipeline processed (e.g., flat fielding, dark subtraction, bad pixel masking).
We registered and combined the images for a given field and filter.
For the resulting mosaics, we derived the astrometric and flux calibrations
with data from 2MASS and UKIDSS. Since $W$ is a custom filter that is
absent from 2MASS and UKIDSS, we calibrated the relative photometry 
among different $W$ images such that their loci of reddened stars in
$W-H$ versus $J-H$ were aligned with each other. Relative photometry
of this kind in $W$ is sufficient for our purposes of identifying late-type
objects based on colors.
We identified sources in the images and measured their aperture photometry
with {\tt starfind} and {\tt phot} in IRAF.
The completeness limits for these data are $J=18.2$, $W=18.0$, and $H=17.5$.
The fields covered by WIRCam are indicated in Figure~\ref{fig:fields}.

\subsection{Color-Magnitude Diagrams}

\begin{figure}[h]
	\centering
	\includegraphics[trim = 0mm 0mm 0mm 0mm, clip=true, scale=.7]{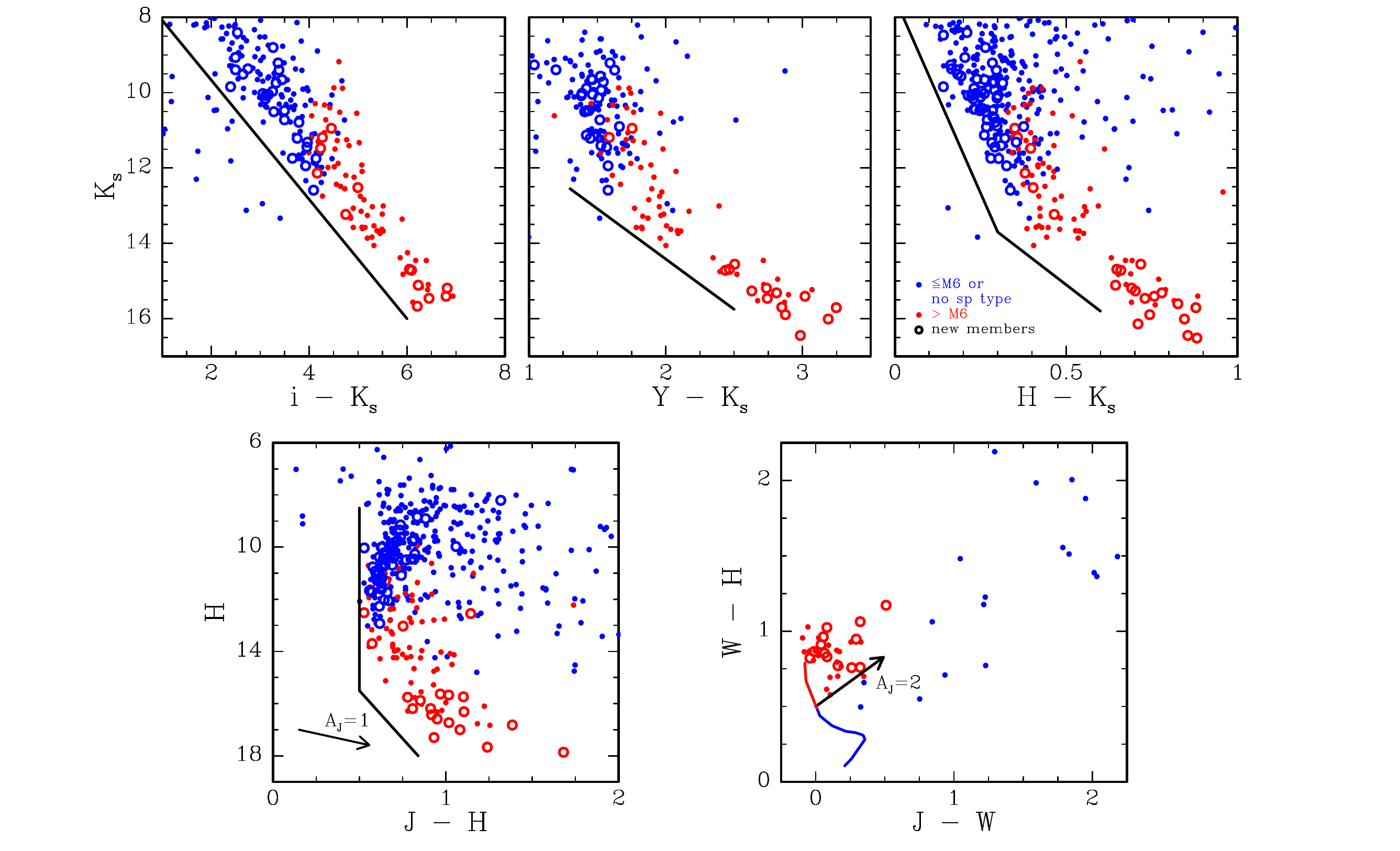}
\caption{
A selection of CMDs and a color-color diagram for previously known members
of Taurus (filled circles) and new members from this work (open circles)
based on photometry from {\it Gaia} DR2, 2MASS, WFCAM (UKIDSS, UHS, new
data), and WIRCam.
From the other stars detected in these surveys, we have selected candidate
members based on positions above the solid boundaries in the CMDs.
The diagram of $W-H$ versus $J-W$ includes the locus of 
young $\leq$M6 and $>$M6 photospheres (blue and red lines)
and is used only for identifying possible
late-type members based on colors above the reddening vector.
The data in the top row of diagrams have been corrected for extinction. 
}
\label{fig:criteria}
\end{figure}

We have identified all matching sources among our new catalogs
and those considered by \citet{esp17}. When multiple measurements in
similar bands were available for a star, we selected the data to adopt
in the manner described by \citet{esp17}.
We omitted photometry with errors $>0.15$~mag in $Y$ and $>0.1$~mag
in the other bands. 
In \citet{esp17}, we constructed diagrams of $K_s$ (or $K$) versus
$G-K_s$, $r-K_s$, $i-K_s$, $z_{\rm P1}-K_s$, $Z-K_s$, $y_{\rm P1}-K_s$,
$Y-K_s$, $H-K_s$, $K_s-[3.6]$, and $K_s-W1$. We also included a diagram
of $W1$ versus $W1-W2$. As explained in that study, we estimated the extinction
for individual stars from $J-H$ and $J-K_s$ and dereddened the photometry
in most of the CMDs. In each CMD, we marked a boundary that followed
the lower envelope of the sequence of known members. Objects
appearing above any of those boundaries and not appearing below any of them
were treated as candidate members. We have applied those CMDs to our updated
compilation of photometry. Four examples of the CMDs are presented in
Figure~\ref{fig:criteria}. In addition, we show a diagram
that makes use of the $W$-band data from WIRCam, $W-H$ versus $J-W$.
The $W$ filter falls within a steam absorption band while $J$ and $H$
encompass continuum on either side of the band, so the combination of 
$J-W$ and $W-H$ can be used to identify late-type objects via their strong
steam absorption. In the diagram of those colors, we have plotted
a reddening vector near the lower edge of the population of known members
later than M6. Objects above that vector are treated as late-type candidates
as long as they are not rejected by any other diagrams. We note that
many of the known $<$M6 members of Taurus within the WIRcam images are
saturated, and hence are absent from the diagram of $W-H$ versus $J-W$.

\subsection{Proper Motions}
\label{sec:pm}

As mentioned in Section~\ref{sec:esp17}, \citet{esp17} measured proper
motions in Taurus with astrometry from 2MASS, PS1, {\it Gaia} DR1, and IRAC.
In addition to those data, we have made use of new motions measured from 
a combination of 2MASS, IRAC, UKIDSS, UHS, and our new WFCAM data, which have
epochs spanning 20 years. The latter four sets of data, which are deeper than
2MASS, span 13 years and reach the lowest masses in Taurus among the
available motions. To measure the proper motions, we began by
aligning each set of astrometry to the {\it Gaia} DR2 reference frame.
Motions were then computed with a linear fit to the available astrometry. 
In Figure~\ref{fig:pm}, we show the resulting motions for individual 
known members of Taurus and for other sources projected against Taurus,
which are represented by density contours. The measurements for the known
members have typical errors of $\sim2$--3~mas~yr$^{-1}$.
Motions with errors of $>10$~mas~yr$^{-1}$ are ignored.
As done with the other catalogs of proper motions
in \citet{esp17}, we have identified candidate members based on motions
that have 1$\sigma$ errors overlapping with a radius of 10~mas yr$^{-1}$
from the median motion of the known members (Figure~\ref{fig:pm}).
Six known members of Taurus do not satisfy this threshold, three of which have
{\it Gaia} DR2 motions that are consistent with membership \citep{luh18}.
The remaining three sources are 2MASS J04355209+2255039, J04354526+2737130,
and J04574903+3015195. The first two objects also have discrepant motions
in {\it Gaia} DR2, but are retained as members for reasons discussed by
\citet{luh18}. The third star is retained as a member since it is 
near known members and is only slightly beyond our proper motion threshold.

\section{Identification of Candidate Members at Stellar Masses}
\label{sec:ident2}

{\it Gaia} DR2 provides high-precision astrometry at an
unprecedented depth for a wide-field survey.
For stars at $G\lesssim20$,
the {\it Gaia} parallaxes and proper motions have typical errors of
$\lesssim0.7$~mas and $\lesssim1.2$~mas~yr$^{-1}$, respectively \citep{gaia18},
which correspond to errors of $\lesssim10$\% and $\lesssim5$\%, respectively,
for unreddened members of Taurus at masses of $\gtrsim0.05$~$M_\odot$.
As a result, {\it Gaia} DR2 enables the precise kinematic identification
of members of Taurus at stellar masses.

\citet{luh18} selected stars from {\it Gaia} DR2 that have 
proper motions and parallaxes that are similar to those of 
the known members of Taurus. In that analysis, the 
parameters {\tt astrometric\_gof\_al} and {\tt astrometric\_excess\_noise} 
from {\it Gaia} DR2 were used to identify stars with poor astrometric fits,
and hence potentially unreliable astrometry.
More recently, \citet{lin18} has presented a
new parameter, RUWE, that serves as a better indicator of the goodness of fit.
He found that the distribution of RUWE in {\it Gaia} DR2 exhibited a break
near 1.4 between the distribution centered at unity expected
for well-behaved fits and a long tail to higher values. 
Thus, RUWE$\lesssim$1.4 could be adopted as a criterion for reliable
astrometry. To illustrate the application of this threshold to Taurus,
we plot in Figure~\ref{fig:ruwe} the distribution of log(RUWE) for known
members (including the new ones from this work) that have parallaxes and
proper motions from {\it Gaia} DR2. We also indicate the subset
of members that are noted in \citet{luh18} and the Appendix
as having discrepant parallaxes (Table~\ref{tab:mem}).
The latter distribution does begin just
above the threshold of 1.4 from \citet{lin18}, supporting its applicability
to Taurus. Above that threshold, the fraction of members with discrepant
astrometry increases with higher values of RUWE.
Most members with RUWE$>1.4$ do not have discrepant astrometry, indicating
that many stars with high RUWE have fits that are sufficiently good for useful
astrometry.

The selection criteria from \citet{luh18} produced a sample of 62 candidate
members of Taurus. Most of the candidates should have spectral types of
M2--M6 based on their colors. In the next section, we present spectroscopic
classifications for 61 of those stars, 54 of which are adopted as members.
The one remaining candidate that lacks a spectrum is {\it Gaia}
157816859599833472.  It is located $6\arcsec$ from a much brighter star, 
HD~30111.  The two stars have similar proper motions, but the parallax of 
HD~30111 ($3.0\pm0.2$~mas) is much smaller than those of Taurus members 
(6--8~mas), so it was not selected as a candidate member.
However, the value of RUWE for HD~30111 is high enough (2.4)
to indicate a poor astrometric fit and potentially unreliable astrometry.
Therefore, based on its proximity
to the candidate {\it Gaia} 157816859599833472 and its similar motion,
we treat HD~30111 as a candidate member.

\citet{luh18} noted three stars that did not satisfy the selection
criteria for candidates but were located within a few arcseconds of
candidates, and hence could be companions to them.
They consist of {\it Gaia} 164475467659453056, {\it Gaia} 3314526695839460352,
and {\it Gaia} 3409647203400743552.
We present spectral classifications for those stars in the next section.

We have searched for additional companions 
that were detected by {\it Gaia} but were not identified as candidate
members by \citet{luh18}.
We began by retrieving all sources from {\it Gaia} DR2 that are within
$5\arcsec$ from known Taurus members. We omitted companions or candidate
companions that were already known and stars that appear below the sequence 
of Taurus members in CMDs of {\it Gaia} photometry. The remaining sample
consists of {\it Gaia} 152416436441091584, {\it Gaia} 3415706130945884416,
{\it Gaia} 148400225409163776, and {\it Gaia} 154586322638884992. 
All of these stars have photometry in only one {\it Gaia} band. The first 
two objects lack measurements of proper motion and parallax.
Those parameters have large uncertainties for 148400225409163776,
but are similar to the measurements for its primary.
The proper motions and parallaxes of 154586322638884992 and its primary
differ significantly, but the latter may have unreliable astrometry
based on the large value of its RUWE (7.2).

The six candidates discussed in this section that lack spectroscopy
are listed in Table~\ref{tab:cand}.

\section{Spectroscopy of Candidate Members}
\label{sec:spec}

\subsection{Observations}
\label{sec:obs}

We have obtained spectra of 140 candidate members of Taurus identified in
Sections~\ref{sec:ident1} and \ref{sec:ident2}\footnote{Three of these
candidates were found independently by \citet{zha18} and are treated
as previously known members in this work.}, three known companions
that lack spectral classifications, and the primary for one of the latter stars.
We also searched for publicly available spectra of our candidates in
the data archives of observatories and spectroscopic surveys, finding
spectra with sufficient signal-to-noise ratios (SNRs) for 38 objects.
Those archival observations consist of 31 optical spectra from the third data
release of the Large Sky Area Multi-Object Fiber Spectroscopic Telescope survey
\citep[LAMOST;][]{cui12,zha12} and seven IR spectra collected through programs
GN-2017A-Q-81, GN-2017B-Q-35 (L. Albert), and GN-2017B-Q-19 (E. Magnier) with
the Gemini Near-Infrared Spectrograph \citep[GNIRS;][]{eli06}.
We present spectra for a total of 168 objects, some of which were
observed with multiple instruments.
This spectroscopic sample includes 61 of the 62 candidates identified
by \citet{luh18}, three additional stars from that study that did not satisfy
the criteria for candidacy but were located very close to candidates
(see Section~\ref{sec:ident2}), three known companions in Taurus that
lack measured spectral types (JH223~B, XEST~20-071~B, V892~Tau~NE),
and XEST~20-071~A, which was observed at the same time as its companion.
The remaining 100 targets in our sample were selected from the candidates
identified in Section~\ref{sec:ident1}. The highest priority was
given to candidates within the area of full $JHK$ coverage from WFCAM
(Figure~\ref{fig:fields}).

We performed our spectroscopy with the Red Channel Spectrograph \citep{sch89}
and the MMT and Magellan Infrared Spectrograph \citep[MMIRS;][]{mcl12} at the
MMT, GNIRS and the Gemini Multi-Object Spectrograph \citep[GMOS;][]{hoo04} at
Gemini North, SpeX \citep{ray03} at the NASA Infrared Telescope Facility (IRTF),
and the Low-Resolution Spectrograph~2 \citep[LRS2;][]{cho14,cho16}
at the Hobby-Eberly Telescope (HET). 
The instrument configurations are summarized in Table~\ref{tab:log}.
The date and instrument for each object are listed in Table \ref{tab:spec}.
The archival data from LAMOST and GNIRS have been included in
Tables~\ref{tab:log} and \ref{tab:spec}.

\begin{figure}[h]
	\centering
	\includegraphics[trim = 0mm 0mm 0mm 0mm, clip=true, scale=.9]{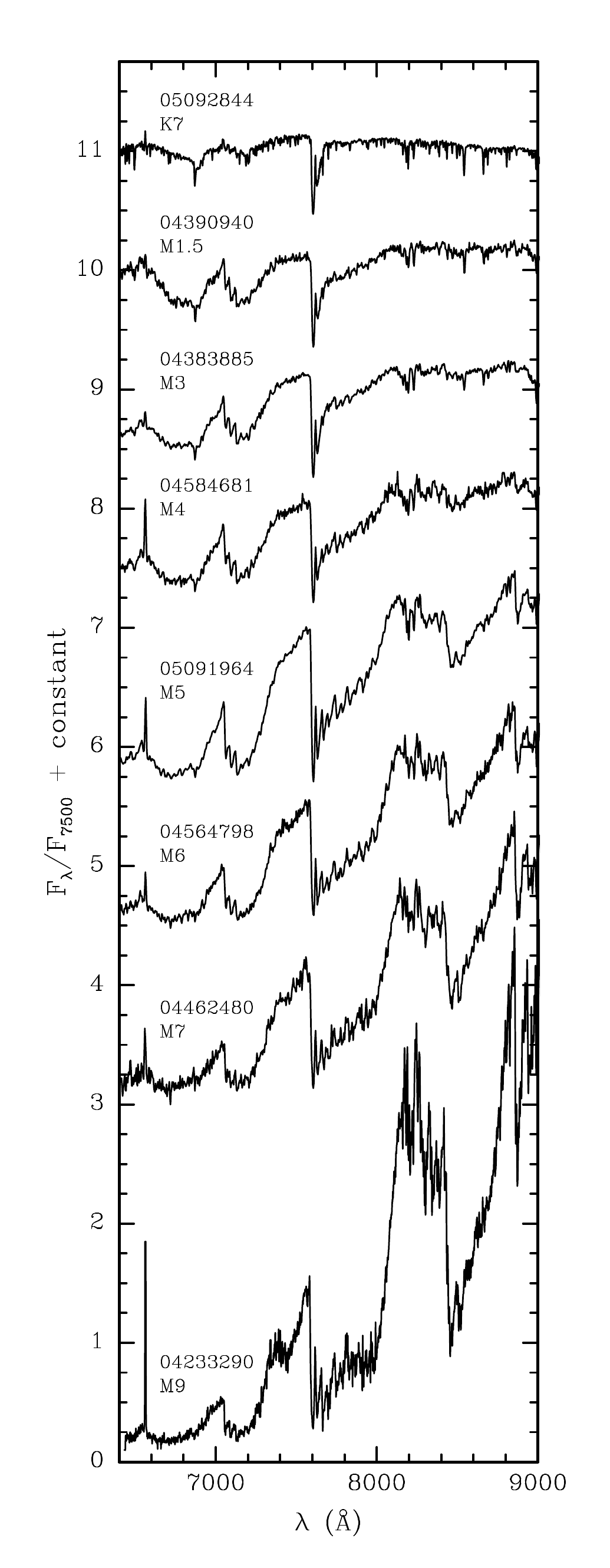}
\caption{
Examples of optical spectra of new Taurus members (Table~\ref{tab:spec}). 
These data are displayed at a resolution of 13 \AA. The data used to create
this figure are available.
}
\label{fig:op}
\end{figure}

We reduced the data from SpeX with the Spextool package \citep{cus04} and 
corrected them for telluric absorption using spectra of A0V stars
\citep{vac03}. The GNIRS and MMIRS data were reduced and corrected for 
telluric absorption in a similar manner using routines within IRAF.
The optical spectra from the Red Channel and GMOS were also reduced with IRAF.
The LRS2 data were processed with the LRS2 Quick-Look
Pipeline (B. L. Indahl, in preparation), which is briefly described by 
\citet{dav18}. Fully reduced spectra are provided by the LAMOST survey.
We present examples of the reduced optical and IR spectra
in Figures \ref{fig:op} and \ref{fig:ir}, respectively. 
All of the reduced spectra are provided in electronic
files that accompany those figures with the exception of the LAMOST
data, which are available from http://www.lamost.org.

\subsection{Spectral Classification} 
\label{sec:specclass}

\begin{figure}[h]
	\centering
	\includegraphics[trim = 0mm 0mm 0mm 0mm, clip=true, scale=.9]{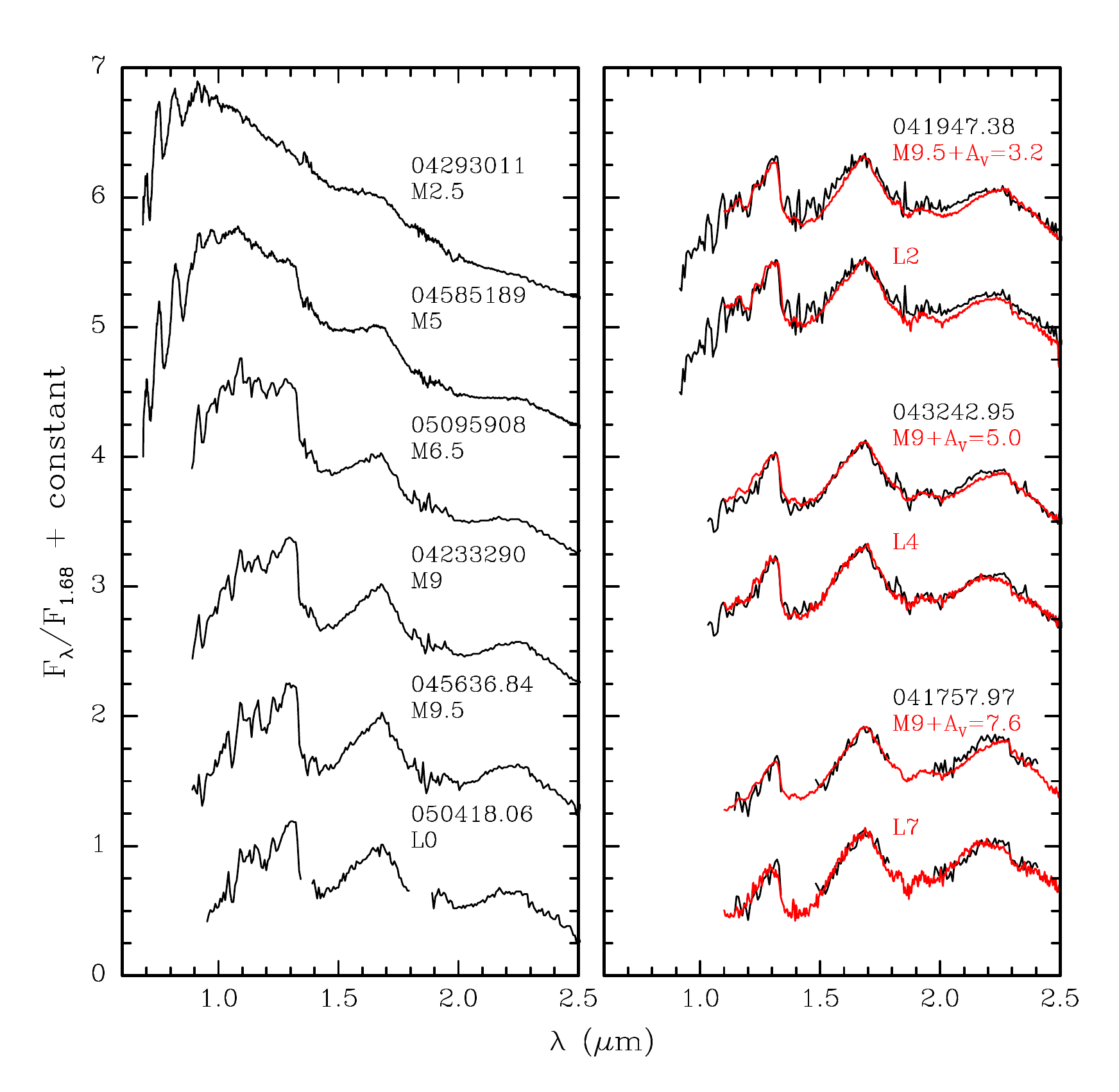}
\caption{
Examples of IR spectra of new Taurus members (Table~\ref{tab:spec}). 
The spectra in the left panel have been dereddened to match the slopes of the
young standards from \citet{luh17}. In the right panel, the observed spectra
of three of the coolest objects are compared to standard spectra that bracket 
their classifications. The data used to create this figure are available.
}
\label{fig:ir}
\end{figure}

\begin{figure}[h]
	\centering
	\includegraphics[trim = 0mm 0mm 0mm 0mm, clip=true, scale=.9]{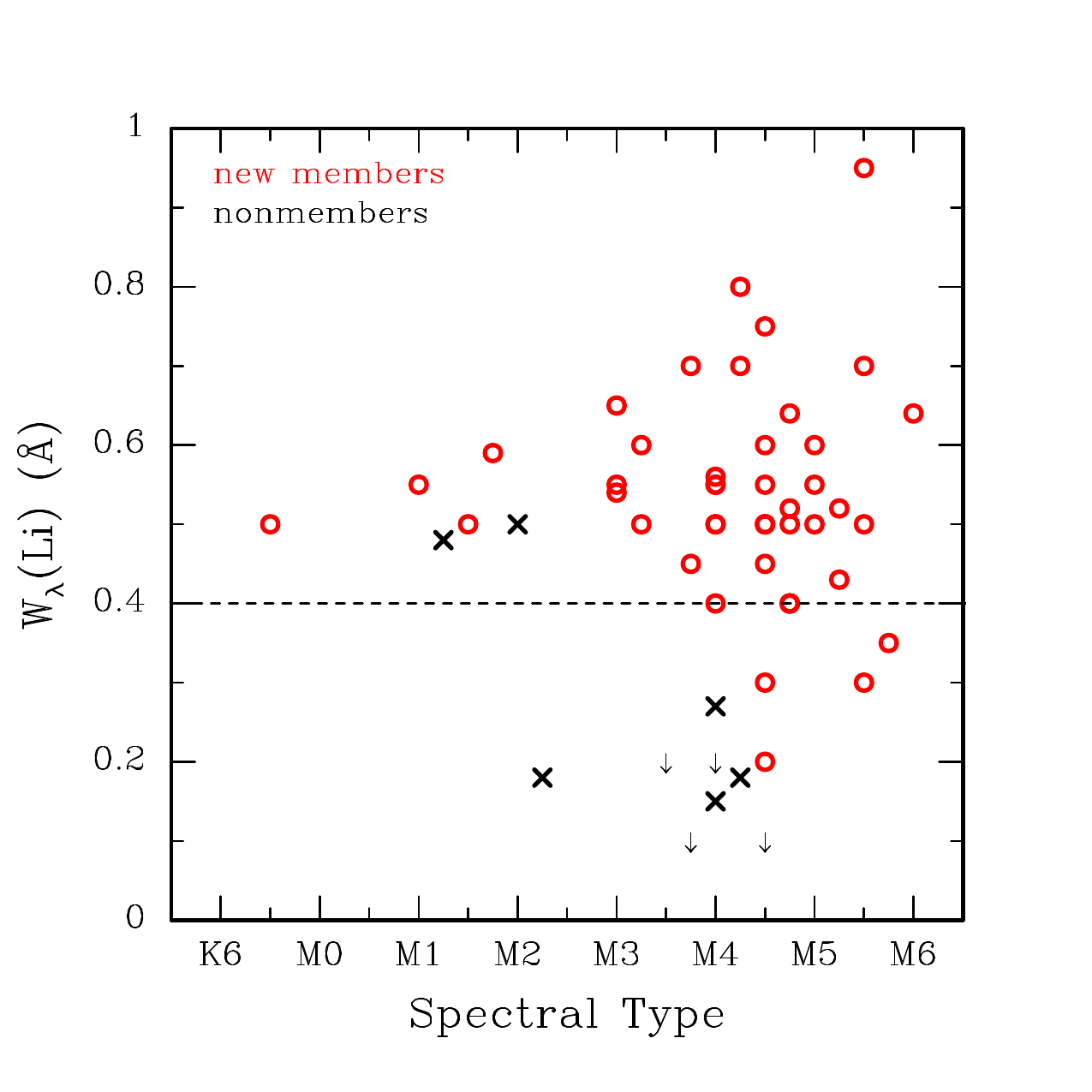}
\caption{
Equivalent widths of Li versus spectral type for candidate members of Taurus.
Known members of Taurus at these types typically have equivalent widths of
$\gtrsim$0.4~\AA\ \citep[dashed line,][]{bas91,mag92,mar94}.
The candidates have been classified as members (circles) or nonmembers (crosses)
based on a combination of their Li strengths and their {\it Gaia} astrometry
(see Section~\ref{sec:specclass} and the Appendix).}
\label{fig:li}
\end{figure}

We have used the spectra from the previous section to estimate spectral types
and to identify evidence of youth that would support membership in Taurus.
Given their colors and magnitudes, the candidates in our spectroscopic sample
should have M/L spectral types if they are members. For these types, we
have utilized diagnostics of youth that include Li {\tt I} absorption at
6707~\AA\ and gravity-sensitive features like the Na {\tt I} doublet near
8190~\AA\ and the shape of the $H$-band continuum \citep{mar96,luh97,luc01}.
Our measurements of the equivalent widths of Li {\tt I} are listed in
Table~\ref{tab:spec} and are plotted versus spectral type
in Figure \ref{fig:li}. For the range of spectral types of the objects with 
useful Li constraints ($\geq$K7), most known members of Taurus have
equivalent widths of $\gtrsim$0.4~\AA\ \citep{bas91,mag92,mar94}.
For young objects at $<$M5 and field dwarfs, we classified the optical spectra
through comparison to dwarf standards \citep{kir91,kir97,hen94}. 
The optical data for young sources at $\geq$M5 were classified with 
the average spectra of dwarf and giant standards \citep{luh97,luh99}.  
For the near-IR spectra, we applied young standards \citep{luh17} and dwarf
standards \citep{cus05,ray09} as appropriate.
The resulting classifications are listed in Table~\ref{tab:spec}.
For the young objects that were observed with IR spectroscopy, we have
used the slopes of those data relative to the best-fitting standards to derive
estimates of extinction. Spectra of young L dwarfs with
low-to-moderate SNRs can be matched by standards across
a wide range of types when extinction is a free parameter \citep{luh17},
as illustrated in Figure~\ref{fig:ir}, where three of the coolest new members
are compared to standard spectra bracketing their classifications.

Moderately young stars ($\sim$10--100~Myr) that are unrelated to the
Taurus clouds \citep[$\sim2$~Myr,][]{pal00}
are scattered across the large field that we have selected
for our survey \citep[][references therein]{luh18}.
As a result, a spectroscopic signature of youth may not be sufficient
evidence of membership in Taurus, particularly if it provides
only a rough constraint on age (e.g., $<100$~Myr).
In addition, some of the young contaminants have proper motions that are
close enough to those of Taurus members that the former can appear to have
motions consistent with membership when the errors are $\gtrsim3$~mas~yr$^{-1}$,
which applies to most non-{\it Gaia} data \citep{luh18}.
Given these considerations, we have taken the following approach to assigning
membership in our spectroscopic sample.
We treat an object as a member if its proper motion and parallax from
{\it Gaia} DR2 support membership (i.e., the candidates from
\citet{luh18}) and its spectrum shows evidence of youth, which is taken
to be $W_\lambda\gtrsim0.4$~\AA\ when a Li measurement is available.
If Li is detected at a weaker level ($\gtrsim$0.15~\AA) and the {\it Gaia}
data agree closely with those of known members, we also adopt the star
as a member (Appendix).
Candidate companions to known members are adopted as members
if they have spectroscopic signatures of youth. Discrepant {\it Gaia}
astrometry is ignored when the astrometric fit is poor (RUWE$\gtrsim1.4$),
which applies to some of the candidate companions.
If {\it Gaia} does not offer reliable measurements of parallax and proper
motion, a candidate is adopted as a member if its available proper motion
data are consistent with membership, its spectrum shows evidence of youth,
it appears within the sequence of known members in CMDs and in a diagram of
$M_K$ versus spectral type, and it is within $\sim1\arcdeg$ of known members.

\begin{figure}[h]
       \centering
       \includegraphics[trim = 0mm 0mm 0mm 0mm, clip=true, scale=.9]{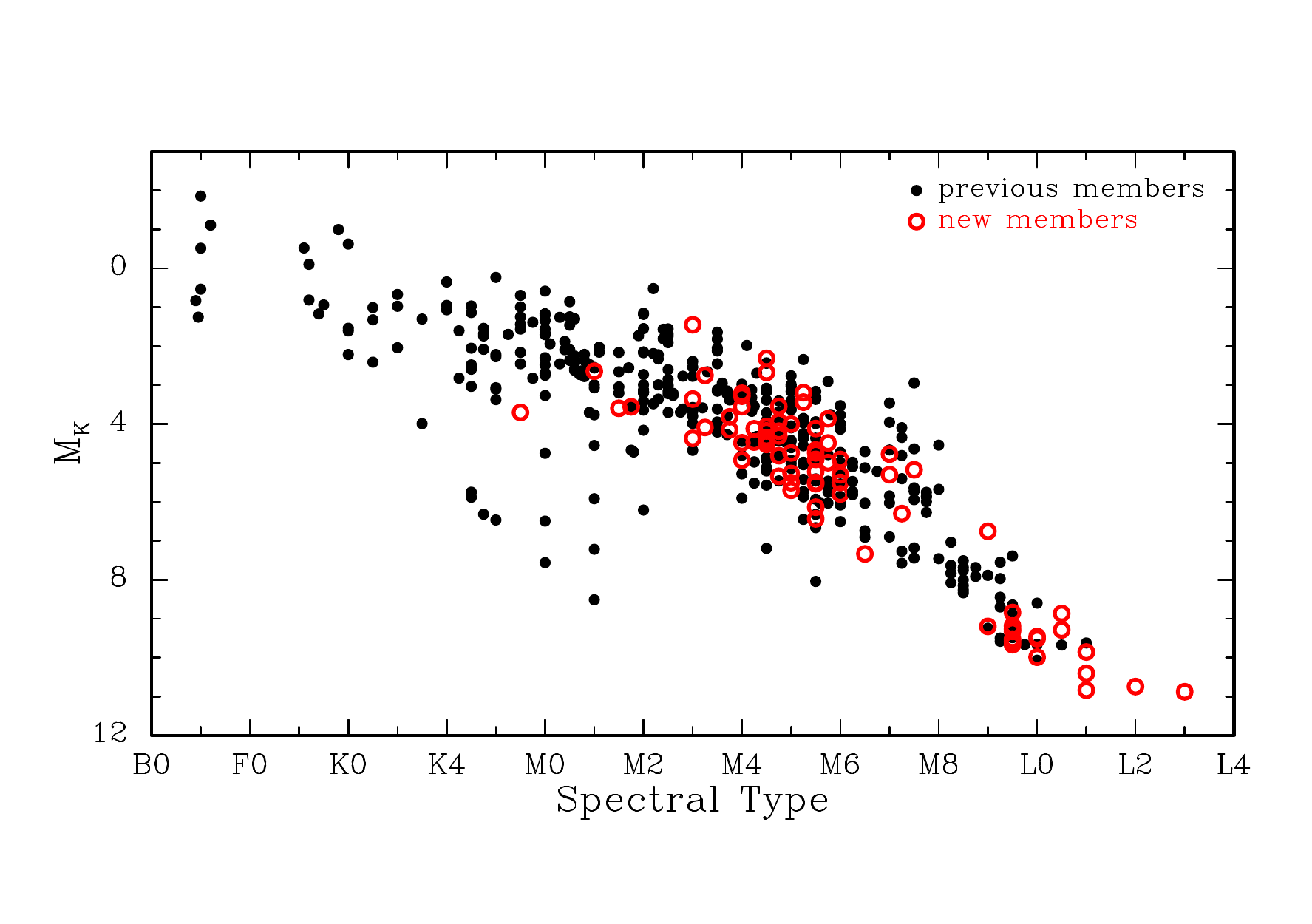}
\caption{
Extinction-corrected $M_K$ versus spectral type for the previously known
members of Taurus (filled circles) and new members from this work
(open circles).
The stars that are below the sequence may be seen primarily in
scattered light, which is plausible given that they have evidence of
circumstellar disks. If a parallax measurement is unavailable for a given
member, we have derived $M_K$ using the median parallax for
the closest Taurus population.
}
\label{fig:mksp}
\end{figure}

Based on the above criteria, 86 of the 168 objects in our spectroscopic sample
are members of Taurus, as indicated in Table~\ref{tab:spec}.
Three members were previously known companions
that lacked spectral classifications, one member is the primary for
one of those companions, and three members were independently found
in a recent survey \citep{zha18}. The remaining 79 members are newly
confirmed in this work. As discussed in Section~\ref{sec:mem}, the
census of Taurus now contains 519 known members. Our survey has
doubled the number of known members at $\geq$M9 and has uncovered the
faintest known members in $M_K$, as illustrated in Figure~\ref{fig:mksp},
where we show extinction-corrected $M_K$ versus spectral type for previously
known members
and our new members. UGCS J041757.97+283233.9 is now the faintest known member.
It has a very red spectrum, which is consistent with spectral types ranging
from M9 ($A_V=7.6$) to L7 ($A_V=0$), as illustrated in Figure~\ref{fig:ir}. 
Assuming a $K$-band bolometric correction for young L dwarfs \citep{fil15},
the median parallax of 7.8~mas for the nearest group of members, and $A_V=3.5$,
we estimate log~$L_{\rm bol}=-3.76$ for UGCS J041757.97+283233.9, which implies
a mass of 0.003--0.01~$M_\odot$ ($\sim$3--10~$M_{\rm Jup}$) for ages of 
1--10~Myr according to evolutionary models \citep{bur97,cha00}.

Most (54/61) of the candidate members identified with {\it Gaia} data
by \citet{luh18} have been adopted as Taurus members.
Among the other candidates that were not part of that sample,
39 are field stars that were observed prior to our WFCAM imaging and
the release of {\it Gaia} DR2 and that would be rejected by our current
criteria that incorporate those data.

\citet{luh18} found that members of Taurus exhibit four distinct populations
in terms of parallax and proper motion, which were given names of
red, blue, green, and cyan.
We have assigned the new members to those populations when the necessary
data are available from {\it Gaia} DR2, as indicated in Table~\ref{tab:mem}
and Figures~\ref{fig:spatial2}--\ref{fig:spatiallast}.

Comments on the spectral types, membership, and kinematics of individual
objects are provided in the Appendix.

\section{Circumstellar Disks}
\label{sec:disk}

\subsection{Disk Detection and Classification}
\label{sec:diskclass}

We have compiled the available mid-IR photometry of the new members of Taurus
from this work to check for evidence of circumstellar disks via the
presence of IR emission in excess above the expected photospheric emission.
We also have performed this analysis on members adopted by \citet{luh18}
that were not examined for disks by \citet{esp17} or earlier studies.
We make use of photometry in $W1$--$W4$ from the AllWISE Source Catalog,
[3.6]--[8.0] from IRAC on {\it Spitzer}, and 
the 24 \micron\ band of the Multiband Imaging Photometer for 
{\it Spitzer} \citep[MIPS;][]{rie04}, which is denoted as [24].
The {\it Spitzer} data were measured in the manner described by \citet{luh10}. 
We present the resulting {\it WISE} and {\it Spitzer} data 
in Table~\ref{tab:mem}. In addition, we have included photometry from those
facilities for previously known members \citep{luh10,esp14,esp17}.

\begin{figure}[h]
	\centering
	\includegraphics[trim = 0mm 0mm 0mm 0mm, clip=true, scale=1]{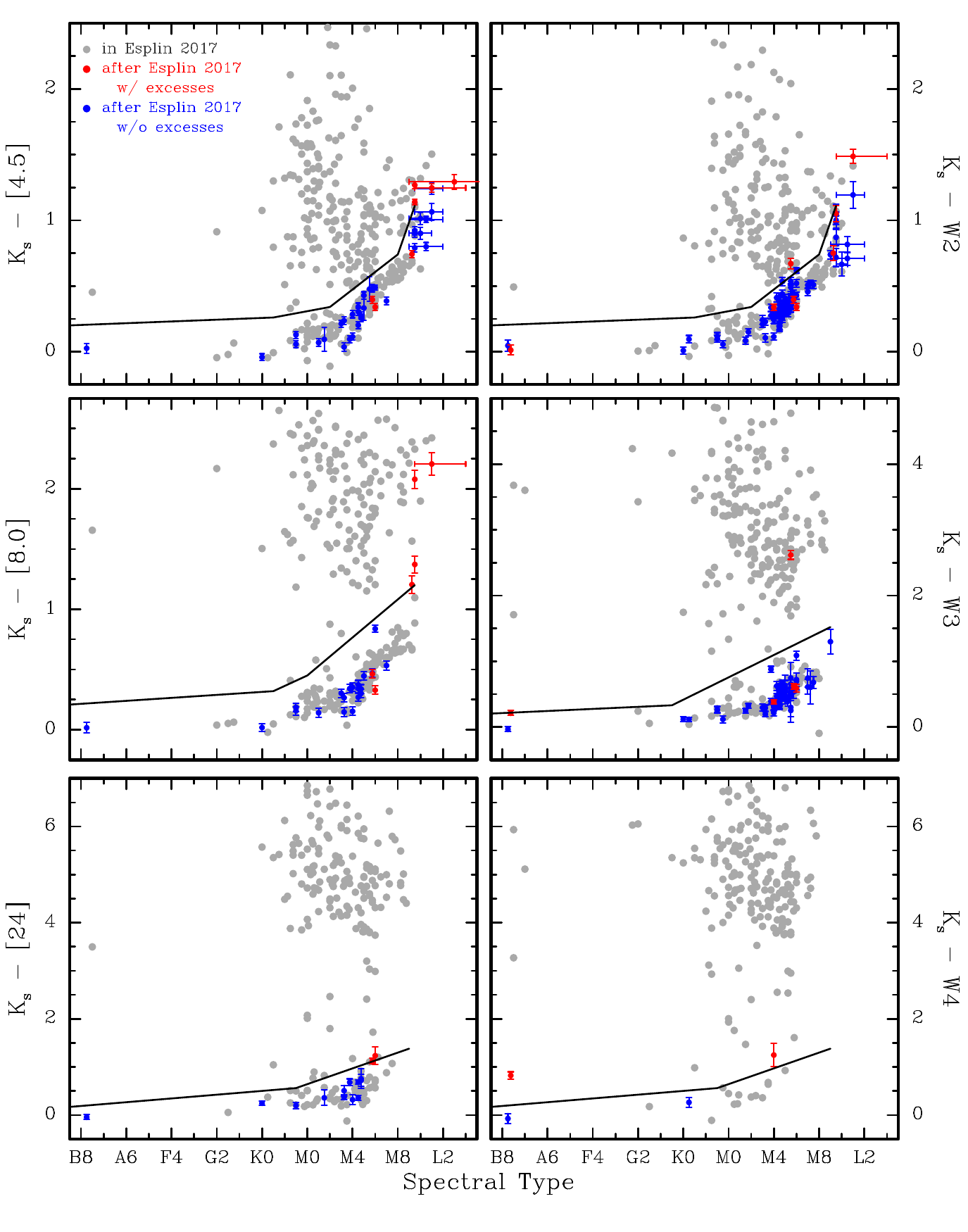}
\caption{
Extinction-corrected IR colors as a function of spectral type for 
known members of Taurus. The members that have been found since \citet{esp17}
are plotted with the errors in their colors and are represented by red and blue
symbols according to presence or absence of excesses in these data.
Excesses have been identified using the indicated thresholds (solid lines),
which were selected to follow the observed photospheric sequence 
in each color \citep{esp14}.
}
\label{fig:excess}
\end{figure}

To determine if excess mid-IR emission is present in a given object
and to classify the evolutionary stage of a detected disk, we have followed
the methods and terminology described in \citet{luh10} and \citet{esp14}
\citep[see also][]{esp18}.  In summary, we calculated extinction-corrected
colors of the mid-IR photometry relative to $K_s$, 
measured color excesses relative to photospheric colors,
and used the sizes of those excesses (when present) to
estimate the evolutionary stages of the disks.
The colors utilized for that analysis are plotted as a function of spectral
type in Figure~\ref{fig:excess} for both previously known members and the
newly classified members. The disks are assigned to the following categories:
optically thick {\it full} (primordial) disks with no large gaps or holes 
that affect the mid-IR spectral energy distribution (SED);
optically thick {\it transitional} disks with large inner holes;
optically thin {\it evolved} disks with no large gaps;
optically thin {\it evolved transitional} disks with large inner holes; and 
optically thin {\it debris} disks that are composed of second-generational
dust from planetesimal collisions \citep{ken05,rie05,her07,luh10,esp12}. 

The Taurus members that have been found since \citet{esp17} are plotted
with red and blue symbols in Figure~\ref{fig:excess} according to the
presence or absence of excess emission, respectively.
Ten of those stars exhibit mid-IR excess emission.
Four of them have excesses only in [24] or $W4$, although one of them,
HD~30378 (B9.5), is only slightly below our adopted excess threshold in $W3$. 
The excesses are small ($\sim0.5$~mag) for the other three stars with excesses
in only [24]/$W4$, which consist of 2MASS J04355694+2351472 (M5.75),
2MASS J04390571+2338112 (M6), and 2MASS J04584681+2954407 (M4).
Two M9.5 members, 2MASS J04213847+2754146 and UGCS J042438.53+264118.5,
have excesses in $[4.5]$, $W2$, and [8.0] and lack reliable detections at
longer wavelengths. 
2MASS J04451654+3141202 (M5.5) has excesses in $W2$ and $W3$.
It is blended with another source in the ${\it WISE}$ images, but it
clearly dominates in $W3$.
2MASS 04282999+2358482 (M9.25) and UGCS J043907.76+264236.0 (M9.5--L4)
have marginal excesses at [8.0] and UGCS J041757.97+283233.9 (M9-L7)
has a marginal excess at [4.5]. 
The latter is the faintest known member in extinction-corrected $K$.
It is difficult to reliably identify the presence of excess emission for the 
coolest members because of the uncertainties in the photospheric colors and
the spectral classifications of young L-type objects \citep{esp17,espetal17}.
The evolutionary stages assigned to the newly identified disks are presented in 
Table~\ref{tab:mem}, where we also include the classifications of previously
known members \citep{luh10,esp14,esp17}.

\subsection{Disk Fraction}

\citet{luh10} measured the fraction of Taurus members that have disks
as a function of spectral type (and hence mass) for known members
observed by {\it Spitzer}.
Since that study, mid-IR photometry has become available for additional
members from both {\it Spitzer} and {\it WISE} and many new members have
been identified. Therefore, it would be worthwhile to perform a new
calculation of the disk fraction in Taurus using our new catalog of members.

The evolutionary stages of young stellar objects consist
of classes 0 and I (protostar+disk+infalling envelope), class II (star+disk),
and class III \citep[star without disk;][]{lad84,lad87,and93,gre94}.
Taurus members that have been previously designated as class 0 or class I
are marked as such in Table~\ref{tab:mem}. We consider all other members
with disks to be class II objects. Some disks have classifications of
``debris/evolved transitional" because we cannot distinguish between these
two classes with the available data. Stars with debris disks are normally
counted as class III objects, but since very few debris disks are expected in
a region as young as Taurus, we treat all of the ``debris/evolved transitional" 
disks as class II.

\begin{figure}[h]
	\centering
	\includegraphics[trim = 0mm 0mm 0mm 0mm, clip=true, scale=1]{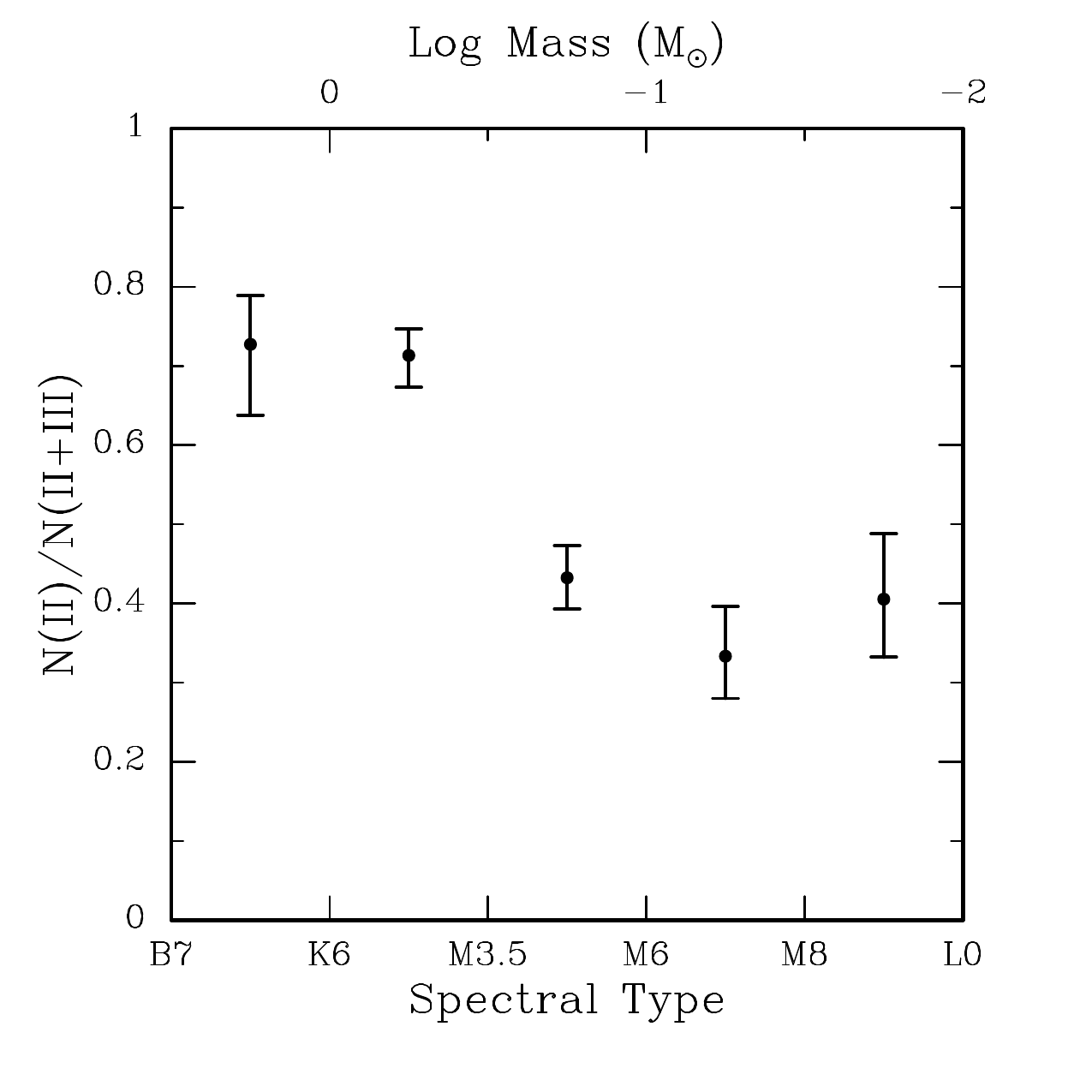}
\caption{
Fraction of Taurus members with circumstellar disks (class II) as a function of 
spectral type (Table~\ref{tab:disks}).
The boundaries of the spectral type bins were chosen to correspond
approximately to logarithmic intervals of mass.
}
\label{fig:disks}
\end{figure}

As done in \citet{luh10}, we define the disk fraction as N(II)/N(II+III)
and we measure it as a function of spectral type using bins of
$<$K6, K6--M3.5, M3.75--M5.75, M6--M8, and M8--M9.75.
We exclude stars that lack measured spectral types, all of which
are protostars or close companions. We also omit objects with spectral types 
of $\geq$L0 because of the difficulty in reliably identifying the presence of
excess emission from disks (Section~\ref{sec:diskclass}).
The resulting disk fraction is tabulated and plotted 
in Table~\ref{tab:disks} and Figure~\ref{fig:disks}, respectively.

The current census of class II members of Taurus should have a high level of
completeness for spectral types earlier than $\sim$M8 \citep{esp14}.
However, the census may be incomplete for class III at high extinctions,
which would lead to an overestimate of the disk fraction when all known
members are considered. To investigate this possibility, we have
computed disk fractions for samples of members that should be complete
for both classes II and III.
Now that most of the candidate members identified with {\it Gaia}
by \citet{luh18} have been observed spectroscopically, 
the census should be complete for both class II and class III
members earlier than M6--M7 at low extinctions \citep[$A_J<1$,][]{luh18}.
Meanwhile, the census within the WFCAM field in Figure~\ref{fig:fields} should
be complete for $\lesssim$L0 at $A_J<1.5$ (Section \ref{sec:imfcomp}).
We find that the disk fractions for $A_J<1$ across the entirety of Taurus
and for $A_J<1.5$ within the WFCAM fields are indistinguishable from 
the disk fraction in Figure~\ref{fig:disks} for all known members.

The disk fraction in Figure~\ref{fig:disks} is near $\sim0.7$ and 0.4 for
spectral types of $\leq$M3.5 and $>$M3.5, respectively. 
A similar trend with spectral type was present in the data from \citet{luh10},
although the disk fraction was slightly higher than
our new measurement ($\sim0.75$ and 0.45).
The disk fraction in Taurus is similar to that in Chamaeleon~I,
which is $\sim$0.7 and 0.45 for $\leq$M3.5 and $>$M3.5 \citep{luh10}. 
For all spectral types combined, Taurus has a disk fraction of $\sim$0.5,
which is roughly similar to the disk fractions of IC~348 and NGC~1333
($\sim$0.4 and 0.6). However, those two clusters
do not show a variation with spectral type \citep{luh16}.

\section{Initial Mass Function}
\label{sec:imf}

\subsection{Completeness}
\label{sec:imfcomp}

To derive constraints on the IMF in Taurus from our new census of members,
we begin by evaluating the completeness of that census.

\begin{figure}[h]
	\centering
	\includegraphics[trim = 0mm 0mm 0mm 0mm, clip=true, scale=1]{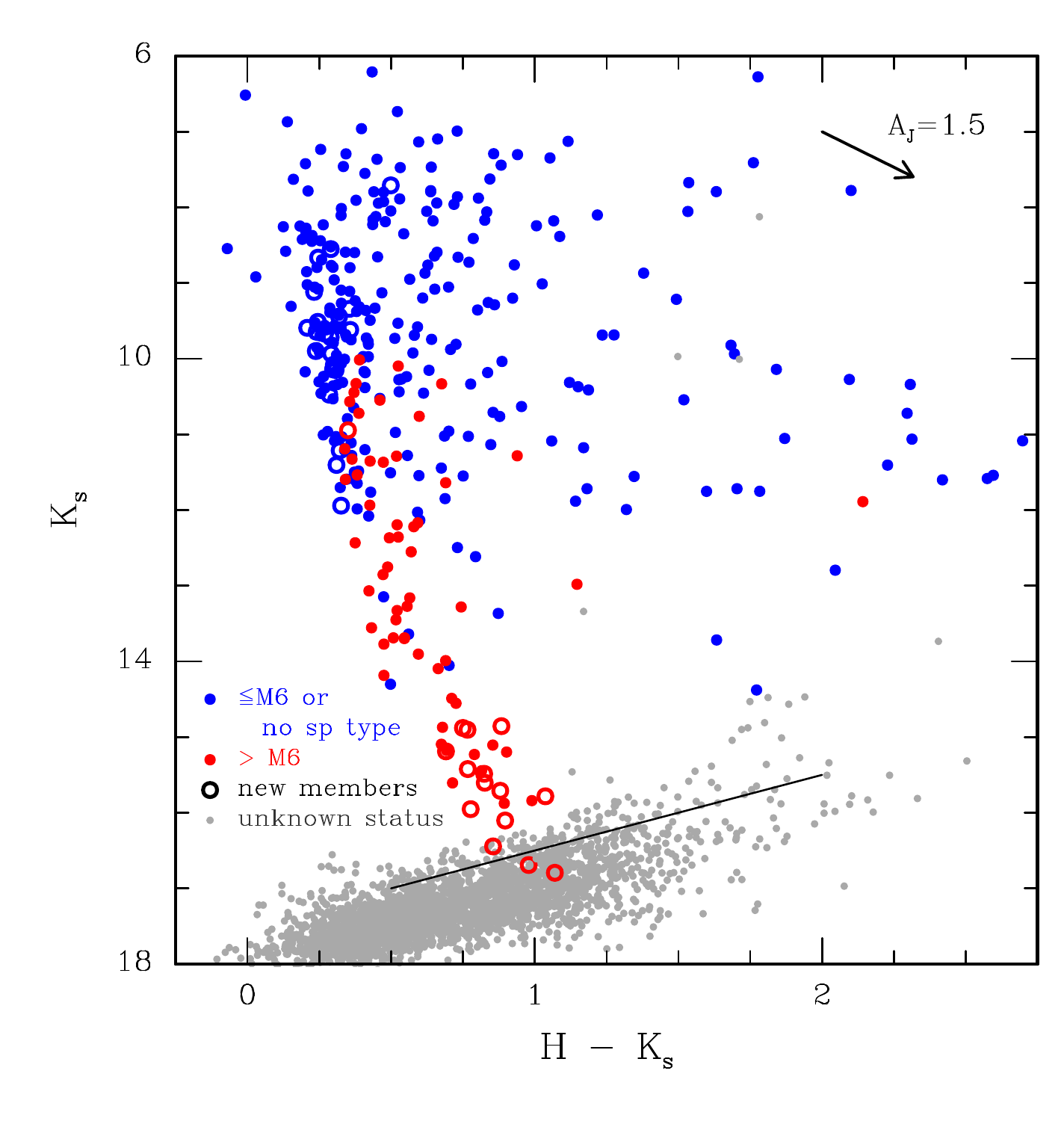}
\caption{
Near-IR CMD of the known members of Taurus within the WFCAM fields
from Figure~\ref{fig:fields} (large filled and open circles)
and the remaining sources in those fields with unconstrained membership 
(small gray points).
The completeness limit of the photometry is indicated (solid line).
}
\label{fig:remain}
\end{figure}

In Section~\ref{sec:ident1}, we focused on the identification of candidate
substellar members within the WFCAM fields in Figure~\ref{fig:fields}.
To evaluate the completeness of our census within that field, 
we employ a CMD constructed from $H$ and $K_s$, which offer the greatest
sensitivity to low-mass members of Taurus among the available bands.
In Figure~\ref{fig:remain}, we plot $K_s$ versus $H-K_s$ for the
known members in the WFCAM fields and all other sources in those fields
that 1) are not rejected by the photometric and proper motion criteria
from Section~\ref{sec:ident1} or the astrometric criteria from \citet{luh18} 
and 2) are not known nonmembers based on spectroscopy or other data.
There are very few remaining sources with undetermined membership status
within a wide range of magnitudes and reddenings.
For instance, the current census within the WFCAM fields
should be complete for an extinction-corrected magnitude of $K_s<15.7$
($\lesssim$L0) for $A_J<1.5$. 

In Section~\ref{sec:ident2}, we adopted the candidate stellar members
that were identified by \citet{luh18} using data for the entirety of Taurus
from {\it Gaia} DR2. That study demonstrated that the census of Taurus
should be complete for spectral types earlier than M6--M7 at $A_J<1$ after
including the {\it Gaia} candidates that are spectroscopically confirmed
to be members.

\subsection{Distributions of Spectral Type, $M_K$, and Mass}

\begin{figure}[h]
	\centering
	\includegraphics[trim = 0mm 0mm 0mm 0mm, clip=true, scale=.8]{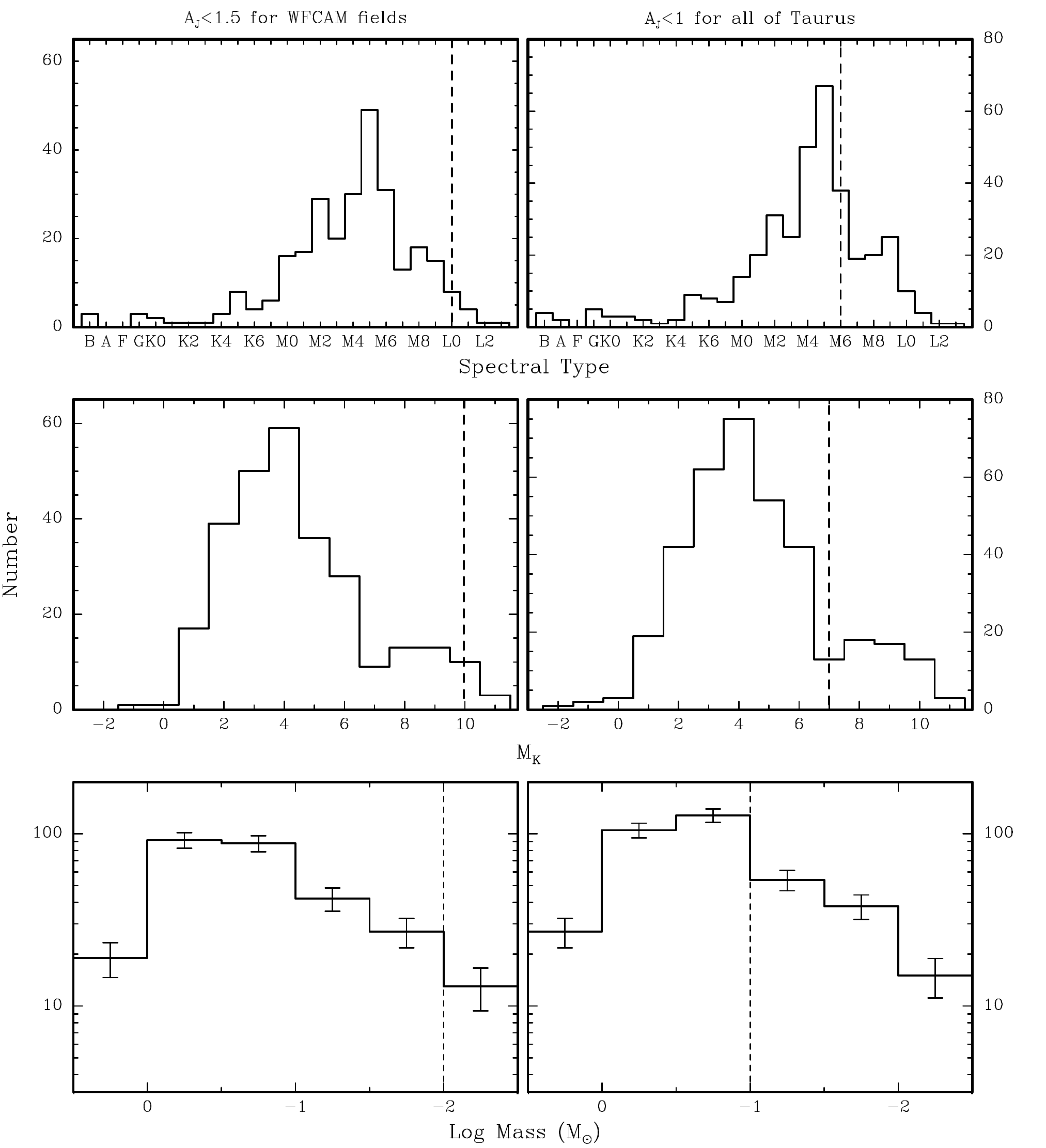}
\caption{
Distributions of spectral types, extinction-corrected $M_K$, and masses
for known members of Taurus with $A_J<1.5$ for the WFCAM fields
from Figure~\ref{fig:fields} (left) and for members with $A_J<1$ in
the entire region (right).
Members were placed in bins of mass based on their spectral types in the
same way done in Figure~\ref{fig:excess}.
The completeness limits for the samples are indicated (dashed lines).
}
\label{fig:imf}
\end{figure}

Based on the analysis in the previous section, we
have defined two extinction-limited samples of known Taurus members that
should have well-defined completeness limits, making them suitable
for characterizing the IMF: members in the WFCAM fields
with $A_J<1.5$ and members at any location with $A_J<1$. 
Stars that lack extinction estimates are excluded from these samples,
which consist of protostars, close companions, and edge-on disks.
As done in our recent studies
of IC~348, NGC~1333, and Chamaeleon~I \citep{luh16,espetal17},
we use distributions of spectral types and extinction-corrected $M_K$
as observational proxies for the IMF.
The distributions of these parameters for our two extinction-limited
samples are presented in Figure~\ref{fig:imf}. 
For objects that lack parallax measurements, we derived $M_K$ using the
median parallax of the nearest population of members.
In addition, we have estimated the IMF for each sample using
distributions of spectral types in which the bins are selected to
approximate logarithmic intervals of mass according to evolutionary models
\citep{bar98,bar15} and the temperature scale for young stars \citep{luh03},
as done for the disk fraction in Figure~\ref{fig:disks}.
The resulting IMFs are shown in Figure~\ref{fig:imf}.

The two samples of Taurus members in Figure~\ref{fig:imf} have similar
distributions, which is not surprising given
the large overlap between the fields in question (i.e., the WFCAM fields
encompass a large majority of known members).
\citet{luh18} found that a $A_J<1$ sample of members
with the {\it Gaia} candidates included exhibited a prominent maximum at M5
($\sim0.15$~$M_\odot$), 
and thus resembled denser clusters like IC~348, NGC~1333, Chamaeleon~I,
and the Orion Nebula Cluster \citep{dar12,hil13,luh16,espetal17}.
Since we have confirmed most of the {\it Gaia}
candidates as members, our $A_J<1$ sample in Figure~\ref{fig:imf} has
a similar distribution of spectral types as in \citet{luh18}.

In the $A_J<1.5$ sample for the WFCAM fields, the distributions of 
spectral type and $M_K$ decrease rapidly
below the peak and remain roughly flat
at substellar masses down to the completeness limit, which
corresponds to $\sim5$--13~$M_{\rm Jup}$ for 
ages of 1--10~Myr according to evolutionary models \citep{bur97,cha00}.
Thus, our completeness limit in that sample does not appear to be near
a low-mass cutoff. The faintest known member, UGCS J041757.97+283233.9, has
an estimated mass of $\sim$3--10~$M_{\rm Jup}$ (Section \ref{sec:specclass}),
which represents an upper limit on the minimum mass in Taurus.
These results are consistent with recent surveys for brown dwarfs
in other star-forming regions \citep{luh16,esp17,zap17,lod18},
young associations \citep{liu13,kel15,sch16b,bes17},
and the solar neighborhood \citep{kir19}, which have found that the
IMF extends down to $\lesssim5$~$M_{\rm Jup}$.
Using the minimum variance unbiased estimator for a power-law distribution,
we calculate a slope of $\alpha=1.0\pm0.1$\footnote{$\alpha$ is defined
such that $dN/dM\propto M^{-\alpha}$. $\alpha=1$ corresponds to a slope
of zero when the mass function is plotted in logarithmic units, as done
in Figure~\ref{fig:imf}.} between
the hydrogen burning limit and the completeness limit (0.01--0.08~$M_\odot$)
in the IMF for the $A_J<1.5$ sample, which is shallower than the slope of 
$\alpha\sim-0.3$ in the lognormal mass function from \citet{cha05}.

\section{Conclusion}

In \citet{esp17}, we searched for substellar members of Taurus using
photometry and proper motions from 2MASS, UKIDSS, PS1, SDSS,
{\it Spitzer}, {\it WISE}, and {\it Gaia} DR1. We have identified additional
candidate members by incorporating new data from UKIRT and CFHT. 
In \citet{luh18}, candidate members at stellar masses were identified
with high-precision proper motions and parallaxes from {\it Gaia} DR2.
We have measured spectral types and assessed membership for
candidates from these two samples using optical and IR spectra.
Through this analysis, we have identified 79 new members of Taurus, which
brings the total number of known members in our census to 519.
Our survey has doubled the number of known members at $\geq$M9 and
has uncovered the faintest known members in $M_K$, which should have
masses extending down to $\sim3$--10~$M_{\rm Jup}$ for ages of 1--10~Myr
\citep{bur97,cha00}.

According to data from {\it Gaia} DR2, our census of Taurus should be nearly
complete for spectral types earlier than M6--M7 at $A_J<1$ across
the entire cloud complex \citep{luh18}.
Meanwhile, we have demonstrated that the census should be
complete for extinction-corrected magnitudes of $K<15.7$ at $A_J<1.5$
within a large field that encompasses $\sim72$\% of the known members.
That magnitude limit corresponds to $\sim5$--13~$M_{\rm Jup}$ for ages of 
1--10~Myr. For the known members within that field and extinction limit,
we have used distributions of spectral types and $M_K$ 
as observational proxies for the IMF. Those distributions remain roughly
constant at substellar masses down to the completeness limit, and thus
show no sign of a decline towards a low-mass cutoff.

We have used mid-IR photometry from {\it Spitzer} and {\it WISE} to search for
evidence of circumstellar disks among the new members from our survey,
as well as the few members adopted by \citet{luh18} that were not examined
for disks by \citet{esp17} or earlier studies.
By combining those results with disk classifications for
all other members in our census \citep{luh10,esp14,esp17}, we have derived
a disk fraction of $\sim$0.7 and 0.4 for spectral types of $\leq$M3.5 and
$>$M3.5, respectively, which is slightly lower than previous measurements
based on less complete catalogs of members \citep{luh10}.

\acknowledgements
K.L. acknowledges support from NASA grant 80NSSC18K0444.
The UKIRT data were obtained through program U/17B/UA05. UKIRT is
owned by the University of Hawaii (UH) and operated by the UH Institute for Astronomy. When the data reported here were acquired, UKIRT was supported by NASA and operated under an agreement among the University of Hawaii, the University of Arizona, and Lockheed Martin Advanced Technology Center; operations were enabled through the cooperation of the East Asian Observatory.
The Gemini data were obtained through
programs GN-2017B-Q-8, GN-2018B-Q-114, GN-2018B-FT-205, GN-2018B-FT-207. 
Gemini is operated by the Association of Universities for Research in Astronomy, Inc., under a cooperative agreement with the NSF on behalf of the Gemini partnership: the National Science Foundation (United States), the National Research Council (Canada), CONICYT (Chile), Ministerio de Ciencia, Tecnolog\'{i}a e Innovaci\'{o}n Productiva (Argentina), and Minist\'{e}rio da Ci\^{e}ncia, Tecnologia e Inova\c{c}\~{a}o (Brazil).
The IRTF is operated by the University of Hawaii under contract NNH14CK55B with NASA.
The MMT Observatory is a joint facility of the University of Arizona and the
Smithsonian Institution.
LRS2 was developed and funded by the University of Texas at Austin McDonald Observatory and Department of Astronomy and by Pennsylvania State University. We thank the Leibniz-Institut f\"ur Astrophysik Potsdam (AIP) and the Institut fur Astrophysik G\"ottingen (IAG) for their contributions to the construction of the integral field units.
The HET is a joint project of the University of Texas at Austin, the
Pennsylvania State University, Stanford University,
Ludwig-Maximillians-Universit\"at M\"unchen, and Georg-August-Universit\"at
G\"ottingen and is named in honor of its principal benefactors, William
P. Hobby and Robert E. Eberly.
The LAMOST data were obtained with the Guoshoujing Telescope,
which is a National Major Scientific Project built by the Chinese Academy of Sciences. Funding for the project has been provided by the National Development and Reform Commission. LAMOST is operated and managed by the National Astronomical Observatories, Chinese Academy of Sciences.
WIRCAM is a joint project of CFHT, Taiwan,
Korea, Canada, France, and the Canada-France-Hawaii Telescope (CFHT) which
is operated by the National Research Council (NRC) of Canada, the Institute
National des Sciences de l'Univers of the Centre National de la Recherche Scientifique of France, and the University of Hawaii.
The {\it Spitzer Space Telescope} and the IPAC Infrared Science Archive (IRSA)
are operated by JPL and Caltech under contract with NASA.
2MASS is a joint project of the University of
Massachusetts and the Infrared Processing and Analysis Center (IPAC) at
Caltech, funded by NASA and the NSF.
Funding for SDSS has been provided by the Alfred P. Sloan Foundation,
the Participating Institutions, the NSF,
the U.S. Department of Energy, NASA, the Japanese Monbukagakusho, the
Max Planck Society, and the Higher Education Funding Council for England.
The SDSS Web Site is http://www.sdss.org/.
The SDSS is managed by the Astrophysical Research Consortium for the
Participating Institutions. The Participating Institutions are the American
Museum of Natural History, Astrophysical Institute Potsdam, University of
Basel, University of Cambridge, Case Western Reserve University, The University
of Chicago, Drexel University, Fermilab, the Institute for Advanced Study, the
Japan Participation Group, The Johns Hopkins University, the Joint Institute
for Nuclear Astrophysics, the Kavli Institute for Particle Astrophysics and
Cosmology, the Korean Scientist Group, the Chinese Academy of Sciences,
Los Alamos National Laboratory, the Max-Planck-Institute for Astronomy,
the Max-Planck-Institute for Astrophysics, New Mexico State University,
Ohio State University, University of Pittsburgh, University of Portsmouth,
Princeton University, the United States Naval Observatory, and the University
of Washington. 
PS1 and its public science archive have been made possible through contributions by the Institute for Astronomy, the University of Hawaii, the Pan-STARRS Project Office, the Max-Planck Society and its participating institutes, the Max Planck Institute for Astronomy, Heidelberg and the Max Planck Institute for Extraterrestrial Physics, Garching, The Johns Hopkins University, Durham University, the University of Edinburgh, the Queen's University Belfast, the Harvard-Smithsonian Center for Astrophysics, the Las Cumbres Observatory Global Telescope Network Incorporated, the National Central University of Taiwan, the Space Telescope Science Institute, the National Aeronautics and Space Administration under Grant NNX08AR22G issued through the Planetary Science Division of the NASA Science Mission Directorate, the NSF Grant AST-1238877, the University of Maryland, Eotvos Lorand University (ELTE), the Los Alamos National Laboratory, and the Gordon and Betty Moore Foundation.
The Center for Exoplanets and Habitable Worlds is supported by the
Pennsylvania State University, the Eberly College of Science, and the
Pennsylvania Space Grant Consortium.

{\it Facilities: } \facility{UKIRT (WFCAM)}, 
\facility{MMT (Red Channel, MMIRS)},
\facility{Gemini:North (GMOS, GNIRS)},
\facility{CFHT (WIRCam)},
\facility{IRTF (SpeX)},
\facility{Spitzer (IRAC)},
\facility{HET (LRS2)}

\appendix

\section{Comments on Individual Sources}

{\it Gaia} 164475467659453056 and 164475467657712256 (2MASS J04161407+2758275~N
and S) comprise a pair with a separation of $0\farcs9$
(Section~\ref{sec:ident2}). 
The first component has a discrepant parallax for Taurus membership
(Figure~\ref{fig:spatial2}), but its astrometry is probably unreliable
based on its large RUWE (5.1) and the {\it Gaia} data for the second
component, which support membership.
LAMOST provides a spectrum of the combined light from the pair.
Since the components have similar $G$ magnitudes, we have assigned the
resulting spectral classification to both of them.
In the diagram of $M_J$ versus spectral type in Figure~\ref{fig:spatial2},
only the second component is plotted, and that is done using half of
the $J$-band flux from 2MASS.

\begin{figure}[h]
	\centering
	\includegraphics[trim = 0mm 0mm 0mm 0mm, clip=true, scale=.7]{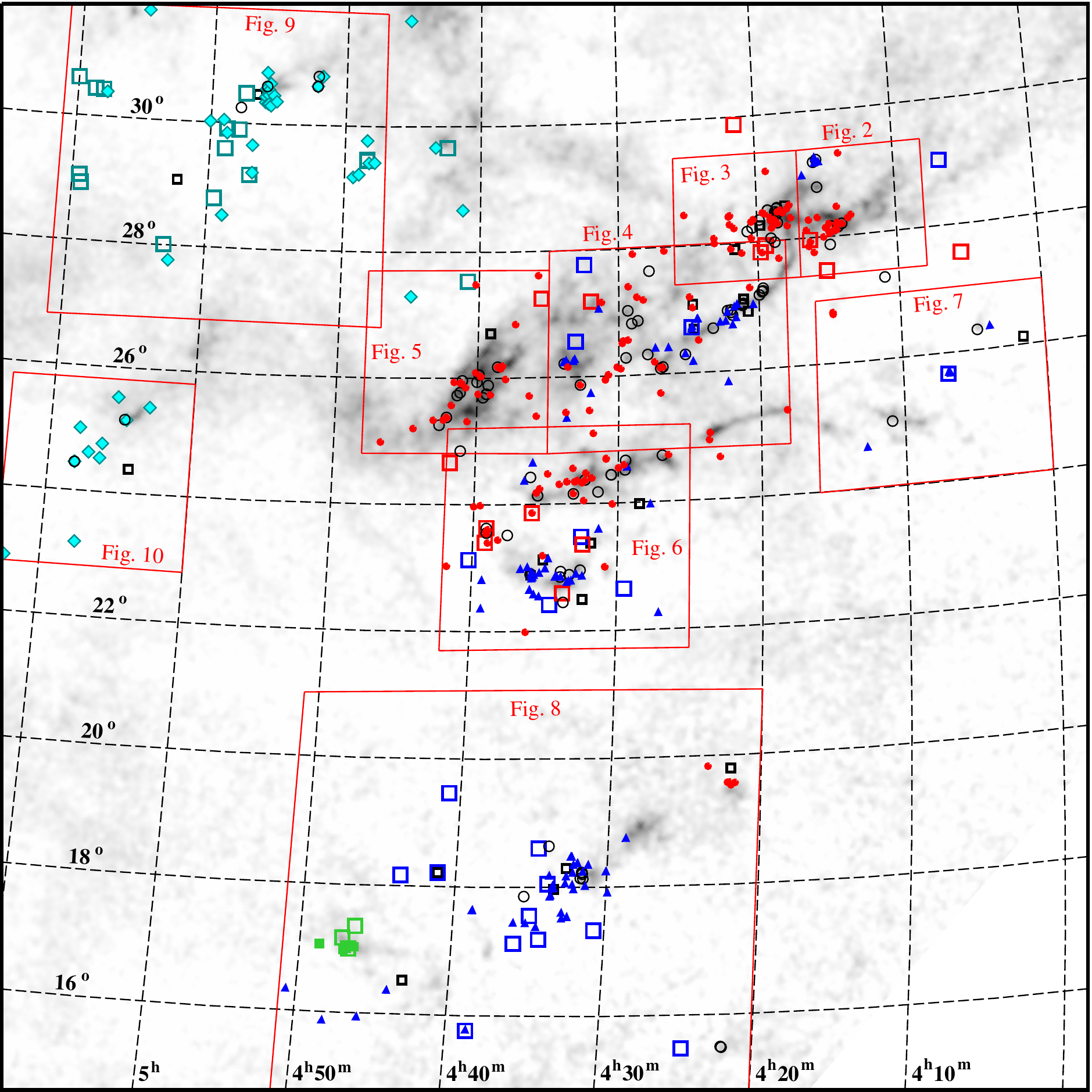}
\caption{
Spatial distribution of the known members of Taurus. 
Previously known members with parallaxes and proper motions from {\it Gaia} 
DR2 are shown with filled symbols (red circles, blue triangles, green squares,
cyan diamonds) and previous members that lack {\it Gaia} data are plotted
with black open circles. The colors of the filled symbols correspond to the
kinematic populations from \citet{luh18}. New members from this work are
plotted with open squares that follow the same color scheme.
The boundaries of the fields encompassed by
Figs.~\ref{fig:spatial2}--\ref{fig:spatiallast} are marked by the red
rectangles. The dark clouds in Taurus are displayed with a map of extinction 
\citep[gray scale;][]{dob05}.
}
\label{fig:spatial1}
\end{figure}

\begin{figure}[h]
	\centering
	\includegraphics[trim = 0mm 0mm 0mm 0mm, clip=true, scale=.7]{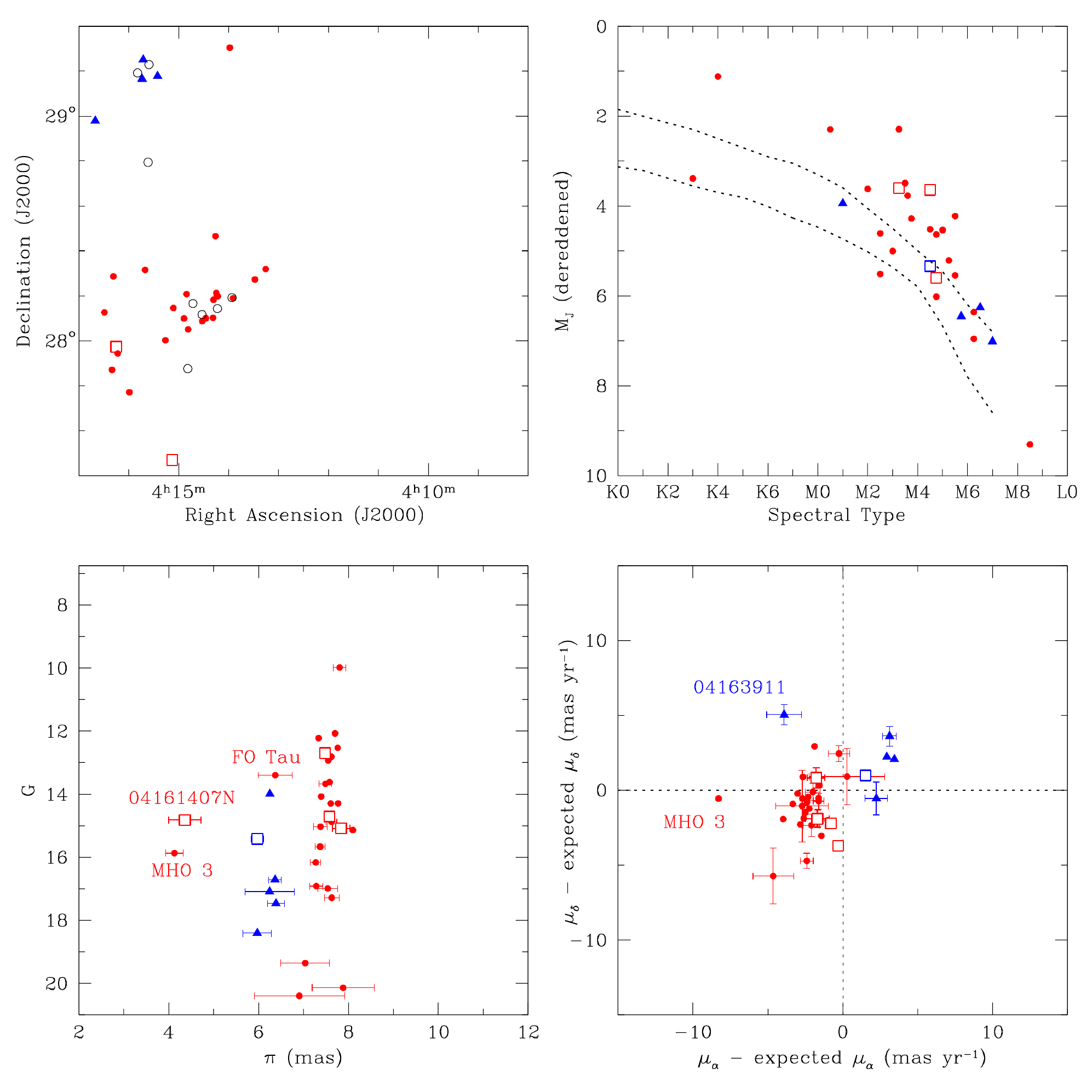}
\caption{
Diagrams of spatial distribution, extinction-corrected $M_J$ versus spectral
type, $G$ versus parallax, and proper motion offsets for previously known
members of Taurus and new members from this work that are projected against 
the B209 cloud. The latter three diagrams consist of stars that have
parallaxes and proper motions from {\it Gaia} DR2.
The symbols are the same as Figure~\ref{fig:spatial1}.
The diagram of $M_J$ versus spectral type includes the median sequences for
Taurus and Upper Sco (upper and lower dotted lines).
The latter has an age of $\sim11$~Myr \citep{pec12,fei16}. 
The proper motion offsets are relative to the values expected for
the positions and parallaxes of the stars assuming the median space velocity
of Taurus members \citep{luh18}.
In the bottom panels, errors are not plotted when they are smaller than
the symbols ($<$0.1 mas, $<$0.5~mas~yr$^{-1}$). 
Stars with discrepant parallaxes and proper motions are labeled
\citep[][Appendix]{luh18}.
}
\label{fig:spatial2}
\end{figure}

\begin{figure}[h]
	\centering
	\includegraphics[trim = 0mm 0mm 0mm 0mm, clip=true, scale=.7]{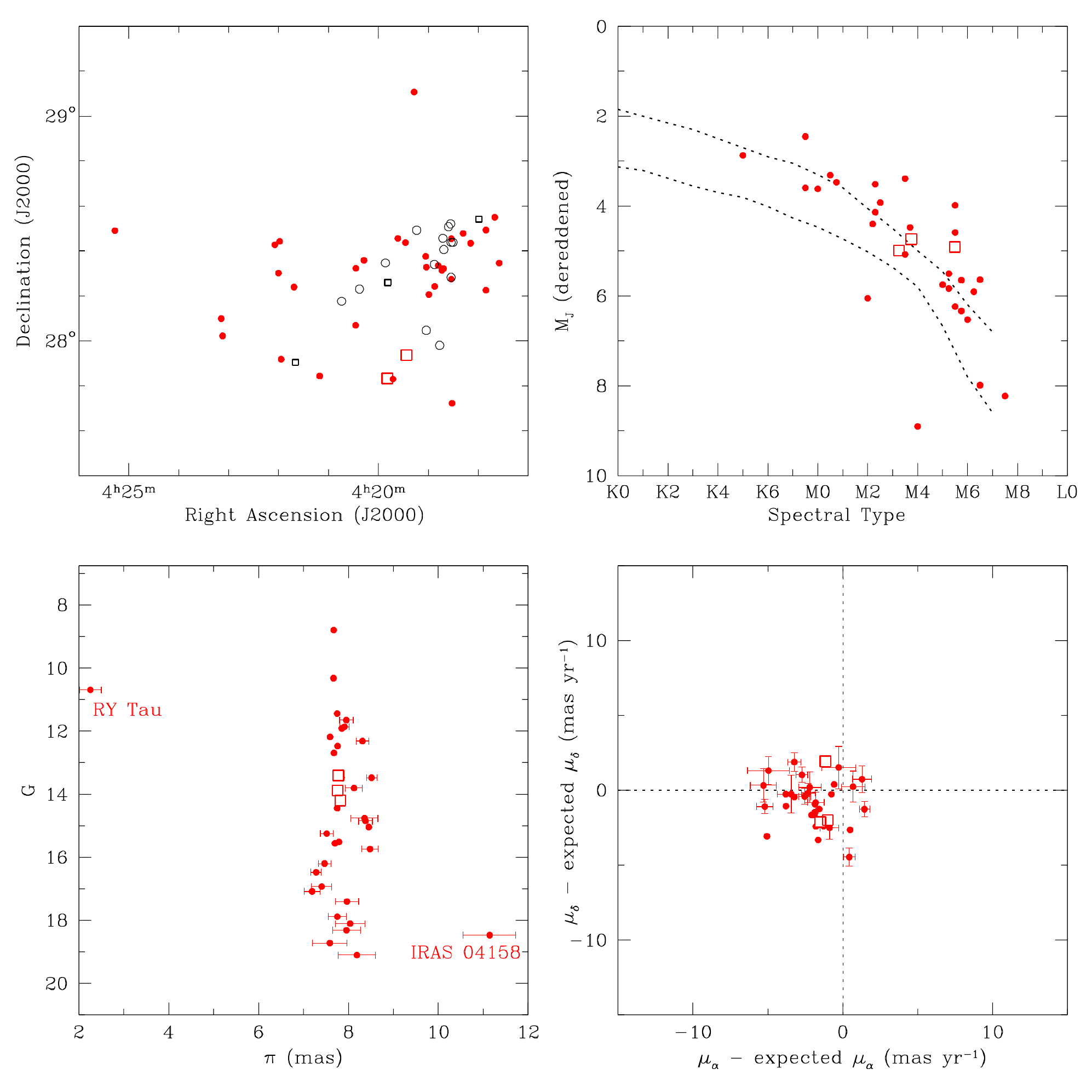}
\caption{
Same as Figure~\ref{fig:spatial2} for members projected against the L1495 cloud.
}
\label{fig:spatial3}
\end{figure}

\begin{figure}[h]
	\centering
	\includegraphics[trim = 0mm 0mm 0mm 0mm, clip=true, scale=.7]{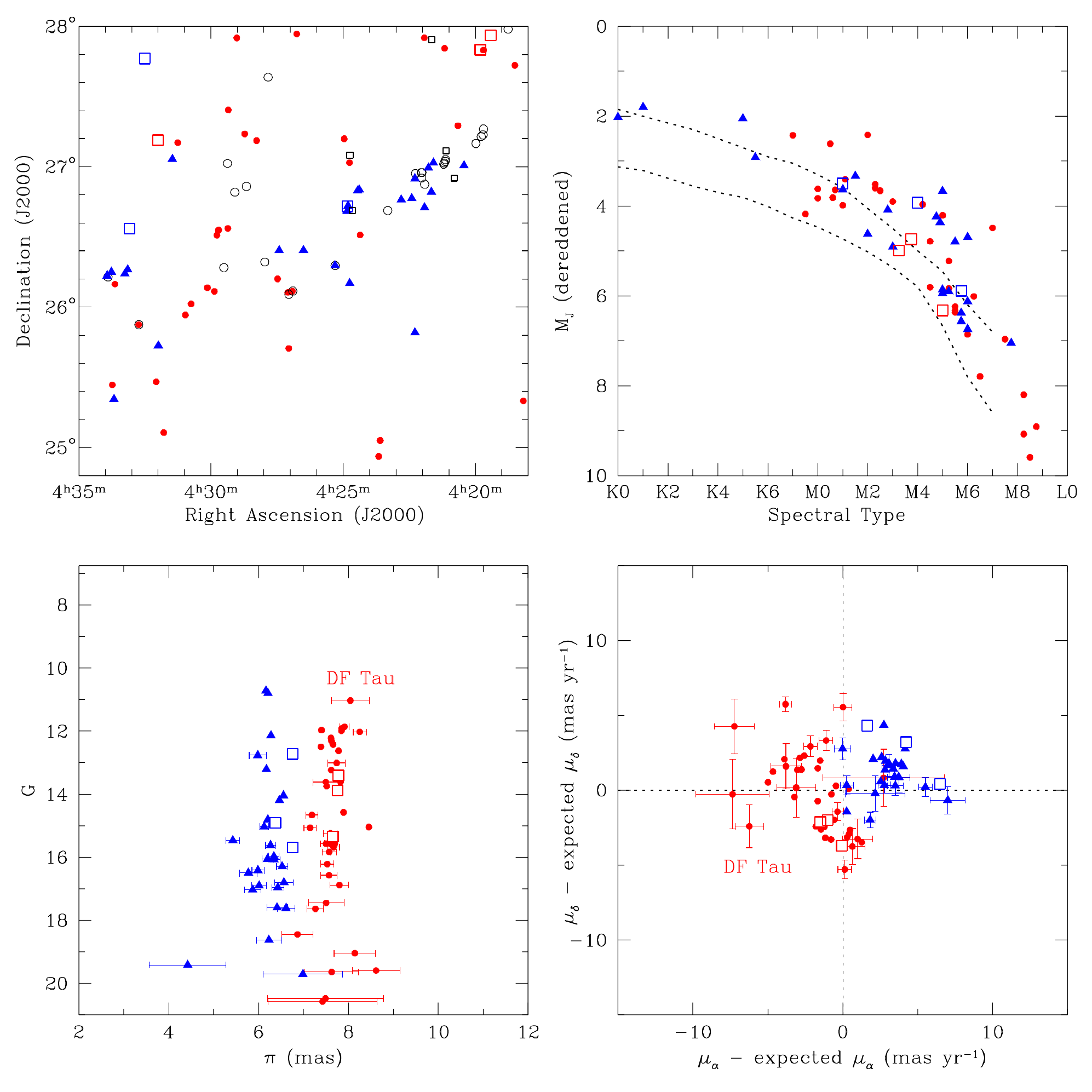}
\caption{
Same as Figure~\ref{fig:spatial2} for members projected against the L1521, B213, and B215 clouds.
}
\label{fig:spatial4}
\end{figure}

\begin{figure}[h]
	\centering
	\includegraphics[trim = 0mm 0mm 0mm 0mm, clip=true, scale=.7]{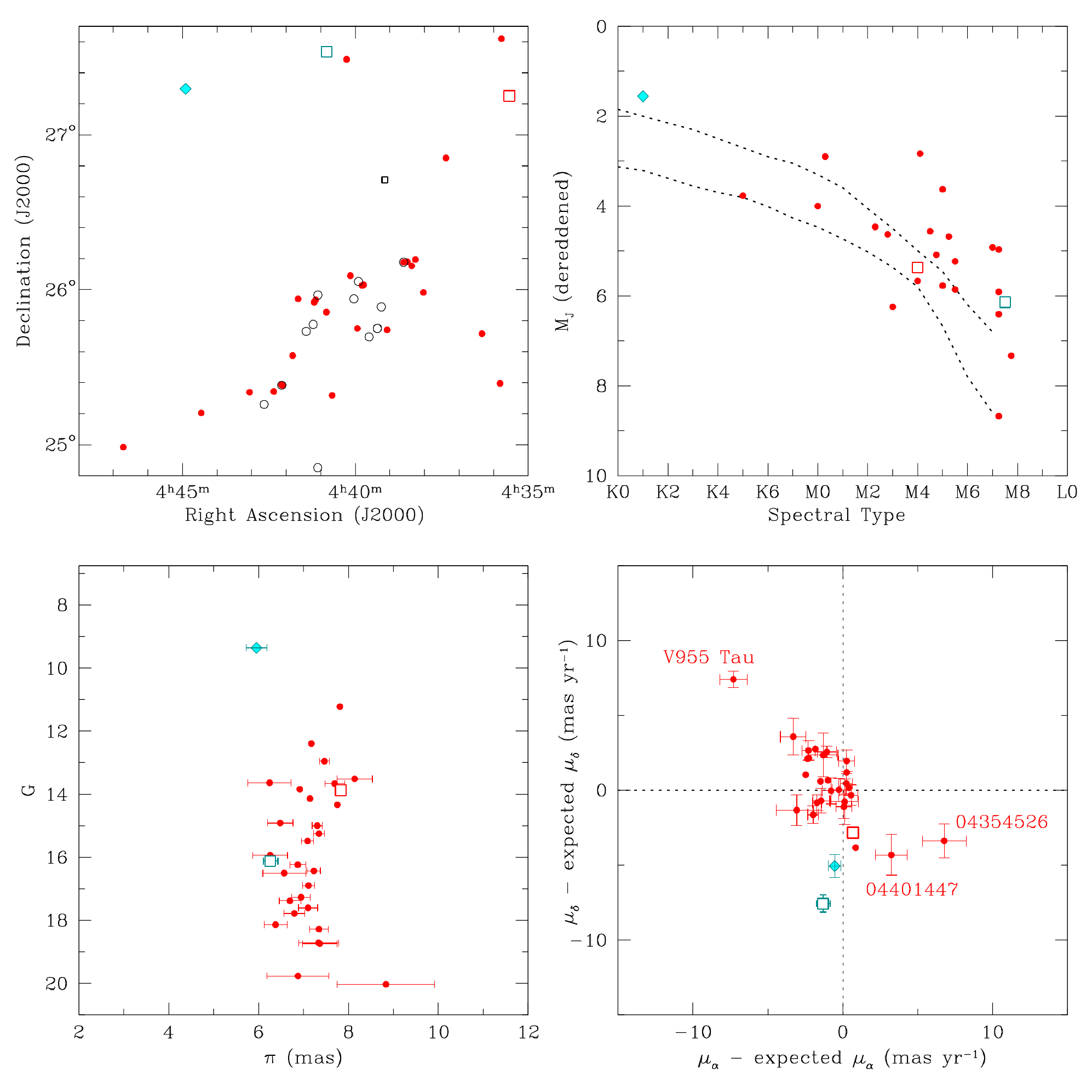}
\caption{
Same as Figure~\ref{fig:spatial2} for members projected against the L1527 cloud.
}
\label{fig:spatial5}
\end{figure}

\begin{figure}[h]
	\centering
	\includegraphics[trim = 0mm 0mm 0mm 0mm, clip=true, scale=.7]{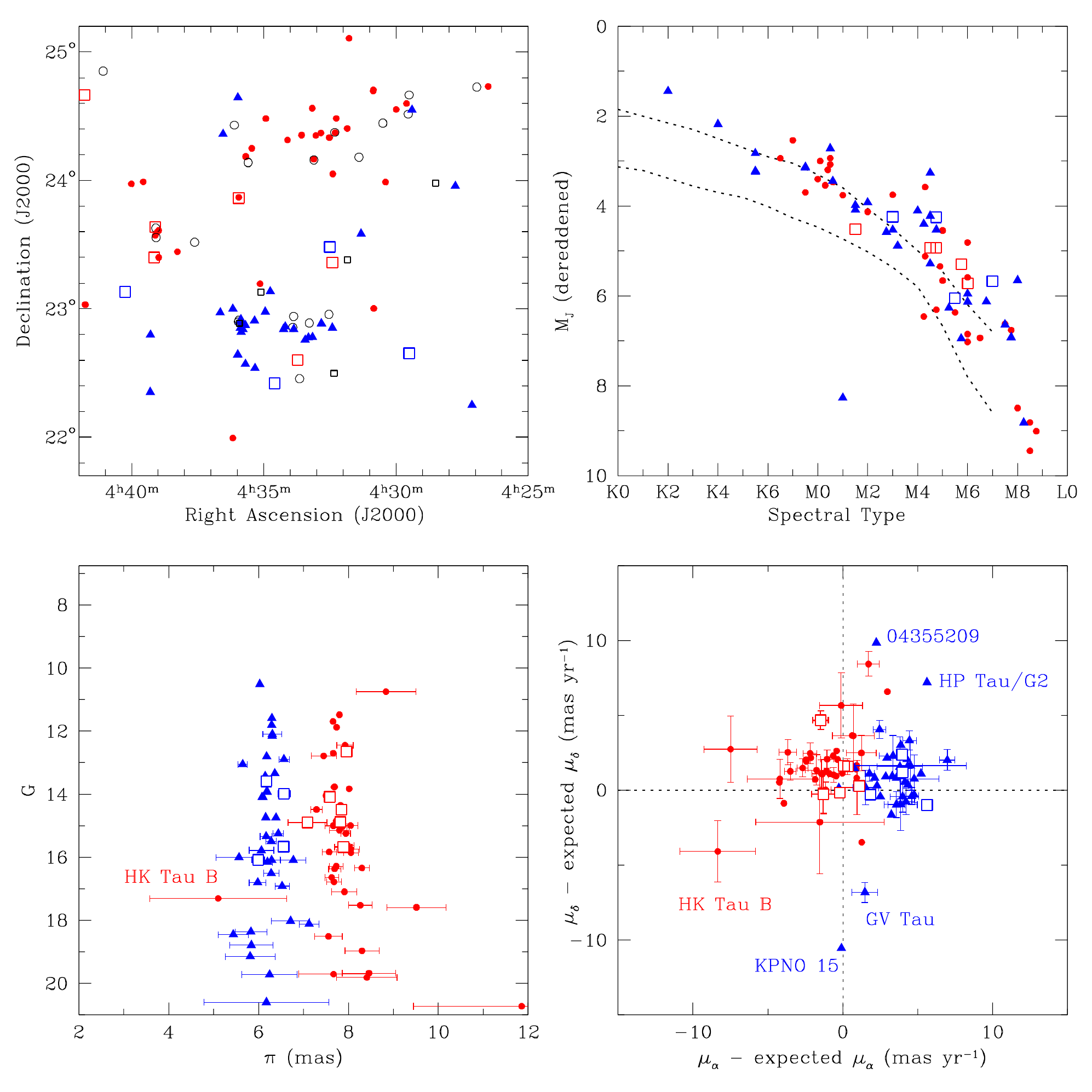}
\caption{
Same as Figure~\ref{fig:spatial2} for members projected against the L1524,
L1529, and L1536 clouds.
}
\label{fig:spatial6}
\end{figure}

\begin{figure}[h]
	\centering
	\includegraphics[trim = 0mm 0mm 0mm 0mm, clip=true, scale=.7]{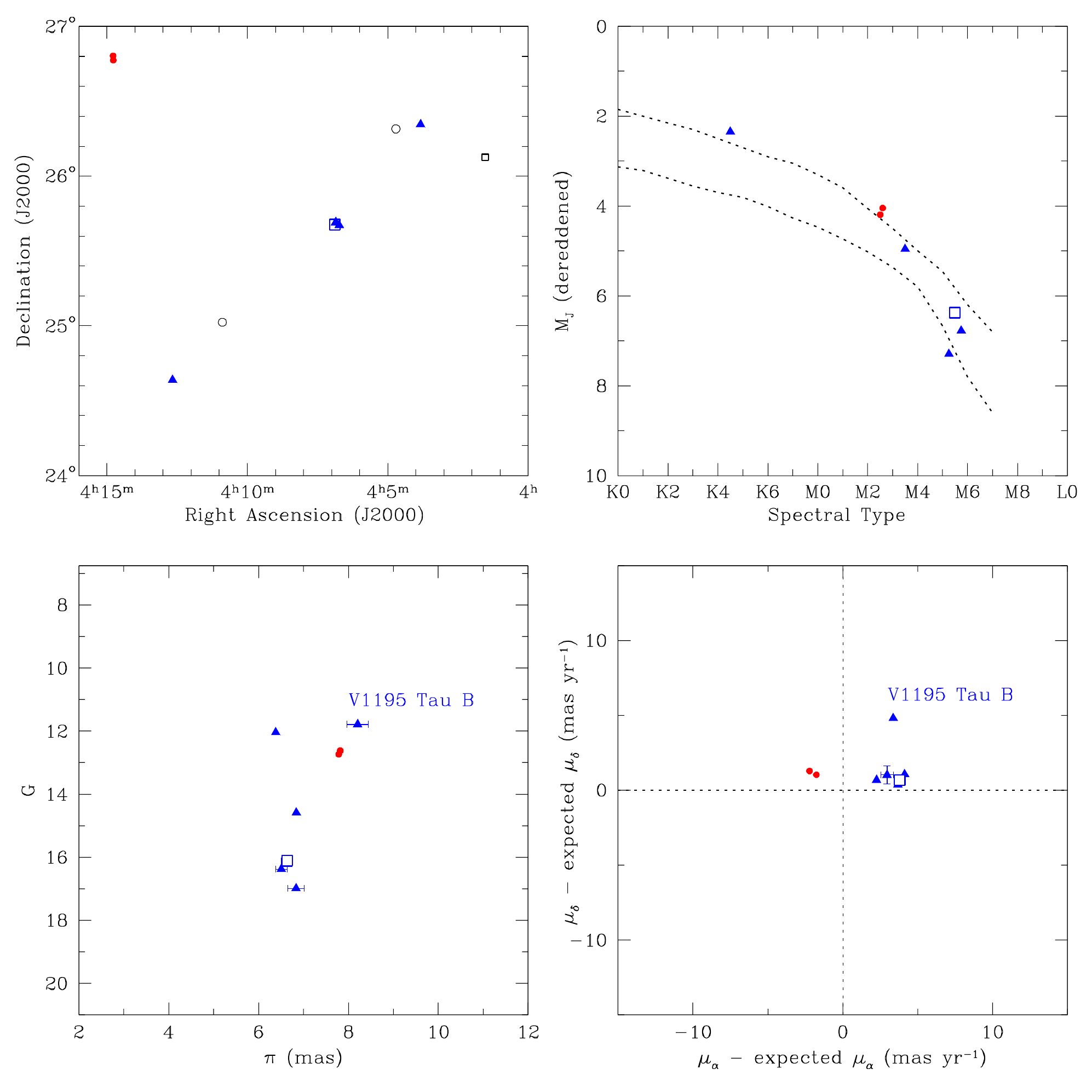}
\caption{
Same as Figure~\ref{fig:spatial2} for members projected against the L1489 and 
L1498 clouds.
}
\label{fig:spatial7}
\end{figure}

\begin{figure}[h]
	\centering
	\includegraphics[trim = 0mm 0mm 0mm 0mm, clip=true, scale=.7]{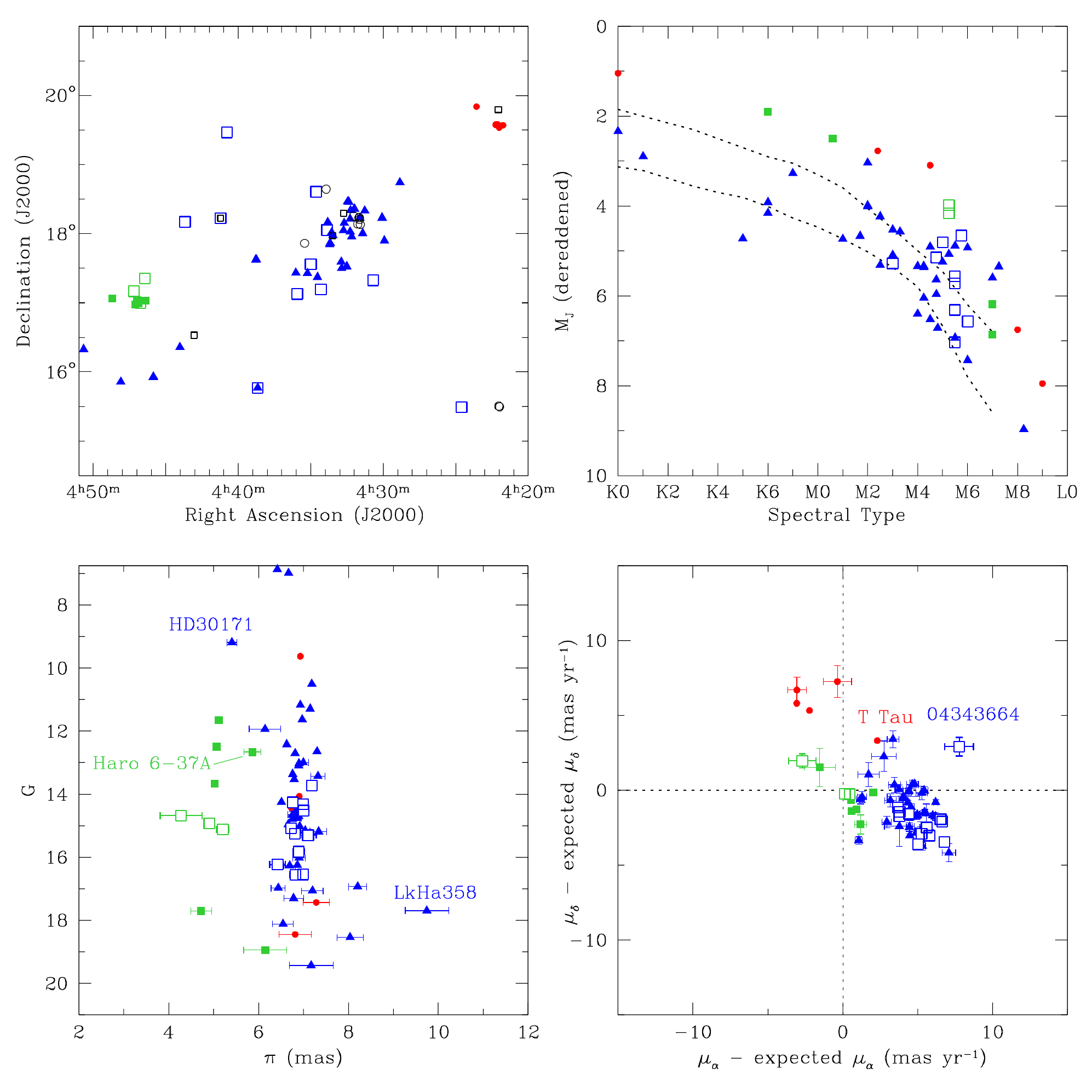}
\caption{
Same as Figure~\ref{fig:spatial2} for members projected against the L1551 and 
L1558 clouds and a small cloud near T Tau.
}
\label{fig:spatial8}
\end{figure}

\begin{figure}[h]
	\centering
	\includegraphics[trim = 0mm 0mm 0mm 0mm, clip=true, scale=.7]{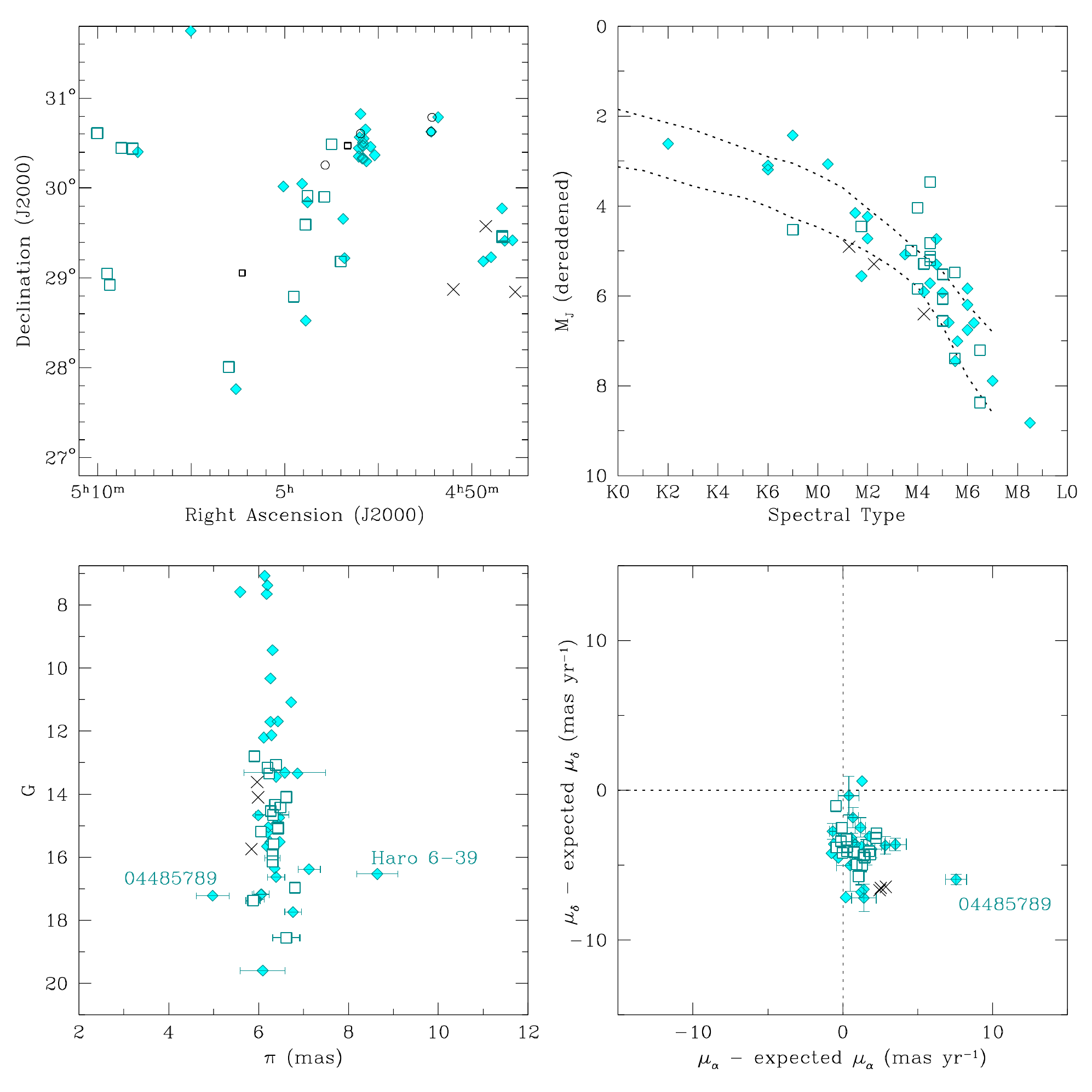}
\caption{
Same as Figure~\ref{fig:spatial2} for members projected against the L1517 cloud.
2MASS J04474012+2850409, 2MASS J04491437+2934354, and 2MASS J04505864+2852218
(crosses) share very similar parallaxes, proper motion offsets, and ages,
indicating that they are likely associated with each other.
Their kinematics are distinct from those of the population of Taurus members
within this field, so they are treated as nonmembers.
}
\label{fig:spatial9}
\end{figure}

\begin{figure}[h]
	\centering
	\includegraphics[trim = 0mm 0mm 0mm 0mm, clip=true, scale=.7]{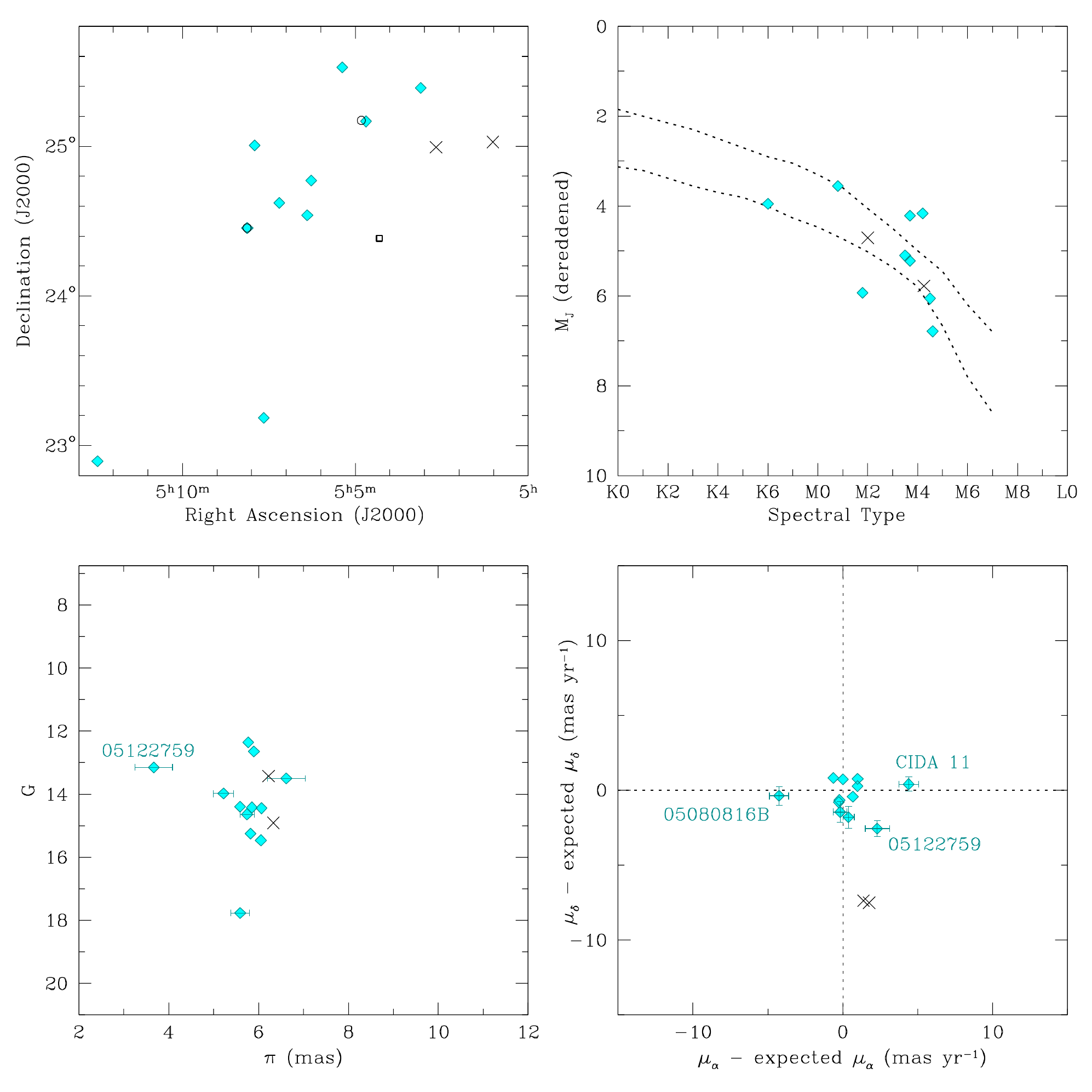}
\caption{
Same as Figure~\ref{fig:spatial2} for members projected against the L1544 cloud.
 2MASS J05010116+2501413 and 2MASS J05023985+2459337 (crosses) share
very similar parallaxes, proper motion offsets, and ages,
indicating that they are likely associated with each other.
Their kinematics are distinct from those of the population of Taurus members
within this field, so they are treated as nonmembers.
}
\label{fig:spatiallast}
\end{figure}

2MASS J04053214+2733139, 2MASS J04064263+2902014, 2MASS J04212650+2952476,
and 2MASS J04422776+2939448 are not within
any of the fields marked in Figure~\ref{fig:spatial1} and shown in
Figures~\ref{fig:spatial2}--\ref{fig:spatiallast}. The data for those
stars are included in the photometric and kinematic diagrams of neighboring
fields, corresponding to Figures~\ref{fig:spatial2}, \ref{fig:spatial2}, 
\ref{fig:spatial3}, and \ref{fig:spatial9}, respectively.

2MASS J04334298+2235566 is intermediate
between the red and blue populations in terms of parallax
(Figure~\ref{fig:spatial6}). We have assigned it to the red group
based on better agreement in proper motion.

The parallax and proper motion of 2MASS J04343664+1836255 are inconsistent
with membership, but those data are probably unreliable (RUWE=8.2). 
In Figure~\ref{fig:spatial8}, the star is beyond the boundaries
of the parallax diagram, labeled in the proper motion diagram, and 
omitted from $M_J$ versus spectral type.

The previously known members 2MASS J05080816+2427150~A and B comprise
a $0\farcs94$ pair. The former lacks measurements of proper motion and
parallax from {\it Gaia}, likely due to a poor astrometric fit (RUWE=71).
The secondary has a discrepant proper motion (Figure~\ref{fig:spatiallast}),
which is probably unreliable (RUWE=3.0).

As mentioned in Section~\ref{sec:mem}, we have adopted
2MASS J04282999+2358482 as a Taurus member, which was classified as a
young M8 dwarf by \citet{giz99}. Using our classification methods, we have
measured a spectral type of M9.25 from the optical spectrum in that study.

\begin{figure}[h]
	\centering
	\includegraphics[trim = 0mm 0mm 0mm 0mm, clip=true, scale=.7]{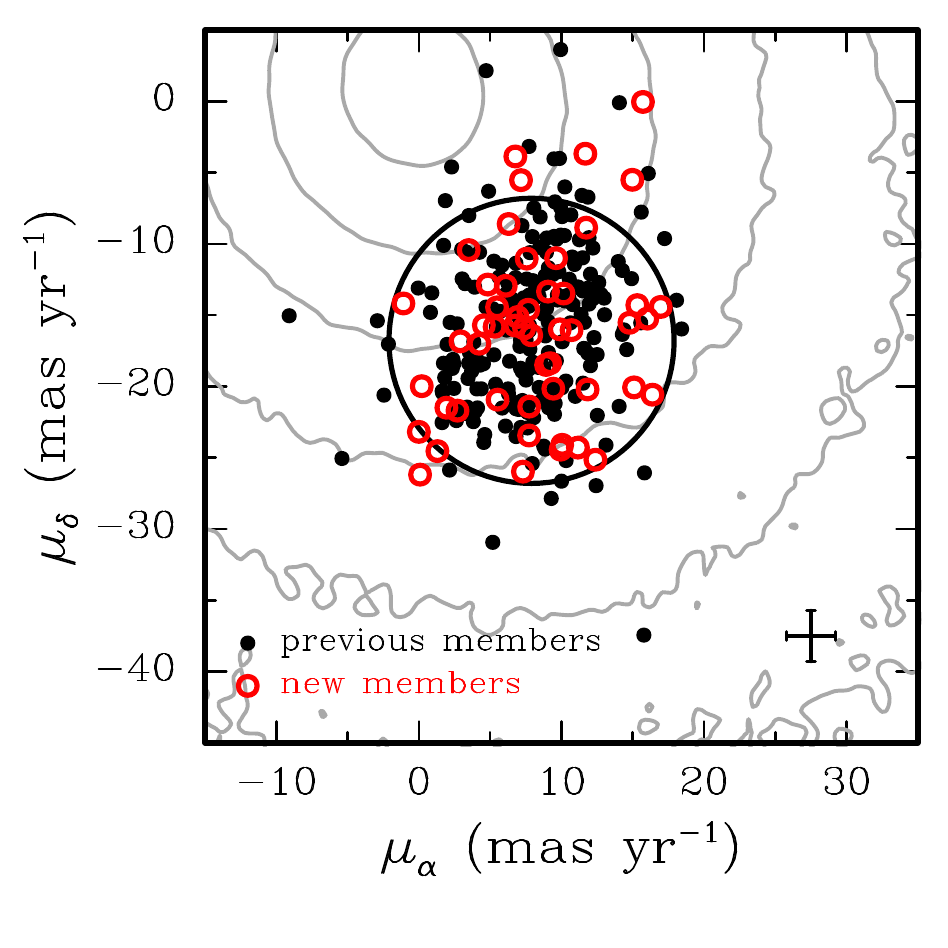}
\caption{
Relative proper motions for known members of Taurus (filled circles) 
and new members from this work (open circles) based on astrometry 
from 2MASS, WFCAM (UKIDSS, UHS, new data), and IRAC.  
Measurements for other sources projected against Taurus are represented by
contours at log(numbers/(mas yr$^{-1}$)$^2$) = 1, 1.5, 2, 2.5, 3, and 3.5.
Sources within 1~$\sigma$ of the large circle are selected as
candidates members. The typical errors for these data are indicated.
}
\label{fig:pm}
\end{figure}

Because of its moderately low SNR, the spectrum
of UGCS J042443.77+270453.4 can be matched with either a young early-L dwarf
or an old mid-L dwarf. We measure a proper motion of
($\mu_\alpha, \mu_\delta=2.9\pm1.3, -16.8\pm5.3$~mas~yr$^{-1}$) from
the data in Section~\ref{sec:pm}, which is consistent with membership 
(Figure~\ref{fig:pm}).
Its small distance from a known member ($3\arcmin$) is also suggestive of
membership. We tentatively adopt it as a member, but a spectrum with higher
SNR would be useful for confirming its youth.

2MASS J04575235+2954072 exhibits evidence of youth in the form of Li
absorption, but the feature is weaker than in most K/M Taurus members 
(0.2~\AA\ vs. $\gtrsim0.4$~\AA). The star agrees well with its neighboring
known members in terms of parallax, proper motion, and the age implied by
$M_J$ versus spectral type, so we adopt it as a member.

The LAMOST spectrum of 2MASS J04053214+2733139 shows Li absorption (0.4~\AA)
while the resolution of the spectrum from Red Channel is too low for a useful
constraint on the feature. The star's parallax and proper motion closely match
those of the red group in Figure~\ref{fig:spatial2}, but given the large
distance from that group ($\sim2\arcdeg$), we consider its membership to be
tentative.

2MASS J04404677+1928033 and 2MASS J05080636+3026233 have moderately weak Li
(0.3~\AA) for Taurus members and share similar kinematics as the blue and cyan
groups in Figures~\ref{fig:spatial8} and \ref{fig:spatial9}, respectively.
We adopt both stars as members, but the membership of the former is considered
tentative because of its large distance from other members.

2MASS J04474012+2850409, 2MASS J04491437+2934354, and 2MASS J04505864+2852218
are among the {\it Gaia} candidates from \citet{luh18} and are located near
each other in the field in Figure~\ref{fig:spatial9}, where they
are plotted as crosses. The second star has Li absorption comparable to
that of Taurus members (0.48~\AA) while the other stars have unusually weak
Li for Taurus (0.18~\AA). The three stars exhibit nearly identical parallaxes,
proper motion offsets, and ages (Figure~\ref{fig:spatial9}),
indicating that they are members of a coeval, comoving group.
The weak Li lines for two of the stars are consistent with the ages
implied by the diagram of $M_J$ versus spectral type (10--20~Myr).
Their parallaxes and motions are distinct from those of the Taurus populations,
so we classify them as nonmembers.
We note that a few of the known members in Figure~\ref{fig:spatial9}
have proper motion offsets that appear to be as discrepant as those of the
three stars in the preceding discussion (e.g., the stars at the top and bottom
of the clump of members). All of those outliers have RUWE$\gtrsim$1.4, so
their discrepant offsets may be due to poor astrometric fits.

2MASS J04404936+2732166 is plotted as the cyan open square
in Figure~\ref{fig:spatial5}. In terms of parallax and proper motion offset,
it matches more closely with the cyan population in the neighboring field
in Figure~\ref{fig:spatial9} than the red population in
Figure~\ref{fig:spatial5}, although there remains a modest difference
in proper motion offsets. Its gravity-sensitive spectral features and
position in the diagram of $M_J$ versus spectral type are consistent with
the age of that cyan population, so we tentatively assign membership to it.

2MASS J04292852+2106069 is located between the fields for
Figures~\ref{fig:spatial6} and \ref{fig:spatial8}.
The presence of Li absorption indicates youth.
Although it was identified as a candidate member based on its parallax and
proper motion \citep{luh18}, it is near the thresholds for selection
in both parameters. Given its remote location and modest discrepancy
in parallax and proper motion relative to other members, we treat its
membership as undetermined.

2MASS J04443916+2224417 shows evidence of moderate youth in its weak
Li absorption (0.15~\AA), but it does not agree well with any of the
groups of known members in terms of its parallax and proper motion,
so we classify it as a nonmember.

2MASS J05023985+2459337 and 2MASS J05010116+2501413 are marked with crosses
in Figure~\ref{fig:spatiallast}. \citet{luh18} adopted the former as a
member of Taurus, but noted that it exhibits a discrepant proper
motion offset relative to Taurus members in its field.
The second star was selected by \citet{luh18} as a candidate member using 
{\it Gaia} astrometry, appearing near the threshold for selection in
proper motion offset. The two stars have similar proper motion offsets,
parallaxes, and ages (Figure~\ref{fig:spatiallast}), indicating that they are 
likely associated with each other. We classify them as nonmembers based 
on their discrepant motions relative to the Taurus members in their field.

2MASS J04195030+2926477 exhibits signatures of youth in its optical
and near-IR spectra. Its {\it Gaia} astrometry is inconsistent with
membership, but those data may not be reliable (RUWE=1.95).
Given its remote location relative to the Taurus groups
(slightly beyond the northern boundary of Figure~\ref{fig:spatial3}),
youth alone is insufficient evidence of membership, so we treat it as
a nonmember.

\begin{figure}[h]
\centering
\includegraphics[trim = 0mm 0mm 0mm 0mm, clip=true, scale=.9]{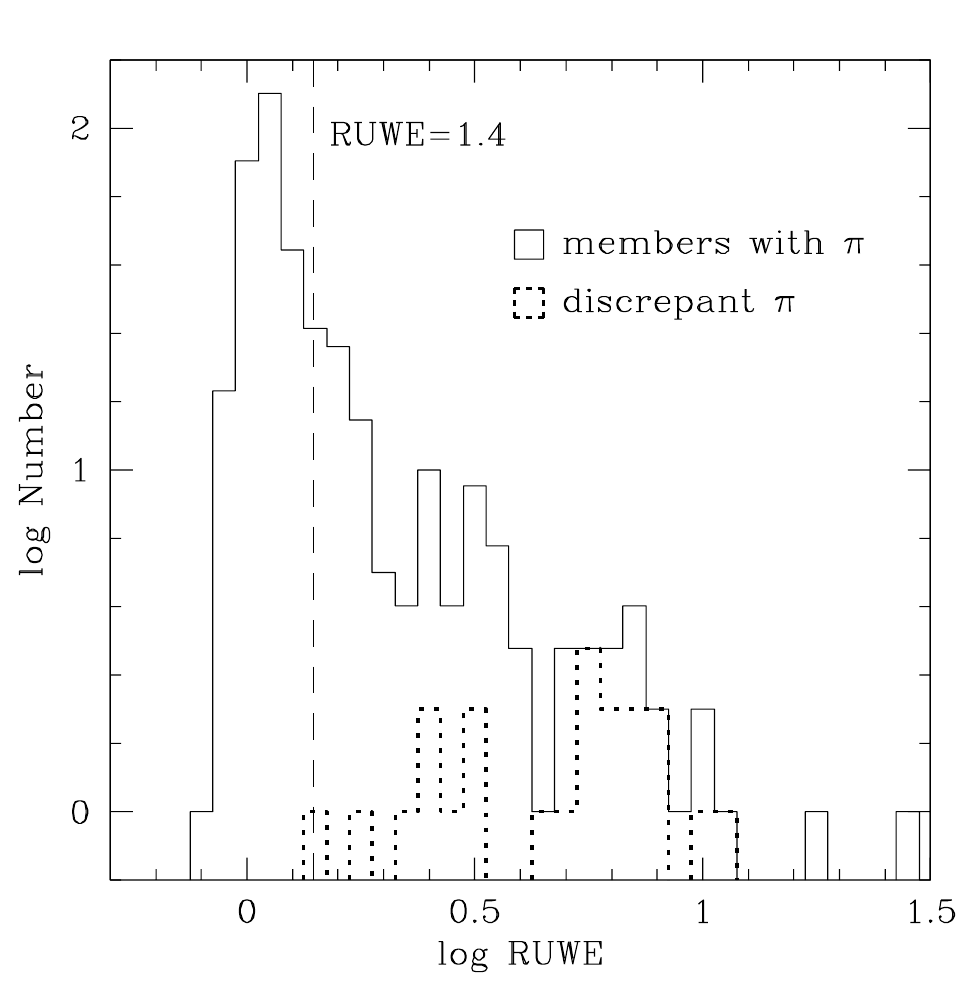}
\caption{
Distributions of log(RUWE) for known members of Taurus with parallax
measurements from {\it Gaia} DR2 (solid histogram)
and the members with discrepant parallaxes 
\citep[dotted histogram,][Appendix]{luh18}. \citet{lin18} suggested
that RUWE$\lesssim$1.4 indicates a good astrometric fit and reliable
astrometry (dashed line).
}
\label{fig:ruwe}
\end{figure}

\citet{luh18} noted that T~Tau differs from the other members
of its red population in terms of its proper motion offset 
(Figure~\ref{fig:spatial8}). Its value of RUWE is somewhat high (1.7,
Figure~\ref{fig:ruwe}), so it may have a poor astrometric fit, which
could explain this discrepancy.

In Section~\ref{sec:obs}, we classified seven of our candidate members of 
Taurus using IR spectra that were publicly available from
Gemini programs GN-2017A-Q-81, GN-2017B-Q-19, and GN-2017B-Q-35.
Although they were not selected as candidates in our analysis, we have
examined the spectra of the remaining 38 objects from those programs.
One of the targets is a previously known member of Taurus,
UGCS J043354.07+225119.1 \citep{esp17}. We classify one source as a galaxy
and 34 sources as field stars. The remaining two objects,
2MASS J04390237+2332133 and UGCS 044344.10+261749.9, appear to be young
and have spectral types of M9--L1 based on comparison to the young standards
from \citet{luh17}. However, both of them fall below the sequence of known
members in multiple CMDs. Young stars that are seen in scattered light
can appear underluminous for their colors, but these two objects do not show
evidence of circumstellar disks in their mid-IR photometry. Instead, they
may be members of the populations of intermediate-age stars ($\gtrsim10$~Myr)
in the direction of Taurus that are unrelated to the cloud complex and its
newly-formed stars \citep{luh18}. We measure proper motions of
$(10.6\pm2.4, -29.0\pm5.3$~mas~yr$^{-1}$) for 2MASS J04390237+2332133
and $(16.4\pm7.4, -45.1\pm2.9$~mas~yr$^{-1}$) for UGCS 044344.10+261749.9.
The former is consistent with membership while the latter is not.
Given its proper motion and proximity to known members, 2MASS J04390237+2332133
seems more likely to be a member, but we treat both objects as nonmembers
based on their underluminous nature.

\clearpage

\LongTables
\begin{deluxetable}{ll}
\tabletypesize{\scriptsize}
\tablewidth{300pt}
\tablecaption{Members of Taurus\label{tab:mem}}
\tablehead{
\colhead{Column Label} &
\colhead{Description}}
\startdata
2MASS & 2MASS Point Source Catalog source name \\
UGCS & UKIDSS Galactic Clusters Survey source name\tablenotemark{a} \\
Gaia & Gaia DR2 source name \\
Name & Other source name \\
RAdeg & Right ascension (J2000) \\
DEdeg & Declination (J2000) \\
Ref-Pos & Reference for right ascension and declination\tablenotemark{b} \\
SpType & Adopted spectral type\tablenotemark{c} \\
GaiapmRA & Proper motion in right ascension from {\it Gaia} DR2\\
e\_GaiapmRA & Error in GaiapmRA \\
GaiapmDec & Proper motion in declination from {\it Gaia} DR2\\
e\_GaiapmDec & Error in GaiapmDec \\
plx & Parallax from {\it Gaia} DR2\\
e\_plx & Error in plx \\
f\_plx & Flag on parallax\tablenotemark{d} \\
Gmag & $G$ magnitude from {\it Gaia} DR2\\
e\_Gmag & Error in Gmag \\
GBPmag & $G_{\rm BP}$ magnitude from {\it Gaia} DR2\\
e\_GBPmag & Error in GBPmag \\
GRPmag & $G_{\rm RP}$ magnitude from {\it Gaia} DR2\\
e\_GRPmag & Error in GRPmag \\
RUWE & re-normalized unit weight error from \citet{lin18} \\
Pop & Population from Luhman (2018) \\
IRpmRA & Proper motion in right ascension from 2MASS/WFCAM/IRAC \\
e\_IRpmRA & Error in IRpmRA \\
IRpmDec & Proper motion in declination from 2MASS/WFCAM/IRAC \\
e\_IRpmDec & Error in IRpmDec \\
Jmag & $J$ magnitude \\
e\_Jmag & Error in Jmag\\
r\_Jmag & Reference for Jmag\tablenotemark{e}\\
Hmag & $H$ magnitude \\
e\_Hmag & Error in Hmag\\
r\_Hmag & Reference for Hmag\tablenotemark{e}\\
Kmag & $K$ or $K_s$ magnitude \\
e\_Kmag & Error in Kmag \\
r\_Kmag & Reference for Kmag\tablenotemark{e}\\
3.8mag  &   {\it Spitzer} [3.6] band magnitude \\
e\_3.6mag &  Error in 3.6mag \\
f\_3.6mag   &  Flag on 3.6mag\tablenotemark{f}\\
4.5mag  &   {\it Spitzer} [4.5] band magnitude \\
e\_4.5mag &  Error in 4.5mag \\
f\_4.5mag   &  Flag on 4.5mag\tablenotemark{f}\\
5.8mag  &   {\it Spitzer} [5.8] band magnitude \\
e\_5.8mag &  Error in 5.8mag \\
f\_5.8mag   &  Flag on 5.8mag\tablenotemark{f}\\
8.0mag  &   {\it Spitzer} [8.0] band magnitude \\
e\_8.0mag &  Error in 8.0mag \\
f\_8.0mag   &  Flag on 8.0mag\tablenotemark{f}\\
24mag  &   {\it Spitzer} [24] band magnitude \\
e\_24mag &  Error in 24mag \\
Ref-Spitz & Reference for Spitzer photometry\tablenotemark{g}\\
f\_24mag   &  Flag on 24mag\tablenotemark{f}\\
W1mag  &   {\it WISE} $W1$ band magnitude\\
e\_W1mag  &  Error in W1mag\\
f\_W1mag  &   Flag on W1mag\tablenotemark{f}\\
W2mag  &   {\it WISE} $W2$ band magnitude\\
e\_W2mag  &  Error in W2mag\\
f\_W2mag  &   Flag on W2mag\tablenotemark{f}\\
W3mag  &   {\it WISE} $W3$ band magnitude\\
e\_W3mag  &  Error in W3mag\\
f\_W3mag  &   Flag on W3mag\tablenotemark{f}\\
W4mag  &   {\it WISE} $W4$ band magnitude\\
e\_W4mag  &  Error in W4mag\\
f\_W4mag  &   Flag on W4mag\tablenotemark{f}\\
Exc4.5  &  Excess present in [4.5]?\\
Exc8.0  &  Excess present in [8.0]?\\
Exc24   &  Excess present in [24]?\\
ExcW2  &  Excess present in $W2$?\\
ExcW3  &  Excess present in $W3$?\\  
ExcW4  &  Excess present in $W4$?\\
DiskType & Disk Type\tablenotemark{h} \\
Aj & Extinction in $J$ \\
f\_Aj & Method for estimating extinction in $J$\tablenotemark{i}
\enddata
\tablenotetext{a}{Based on coordinates from Data Release 10 of the UKIDSS
Galactic Clusters Survey for stars with $K_s > 10$ from 2MASS.}
\tablenotetext{b}{Sources of the right ascension and declination are 
are the 2MASS Point Source Catalog, Gaia DR2, UKIDSS Data Release 10,
and images from the Spitzer Space Telescope \citep{luh10}.}
\tablenotetext{c}{Spectral types adopted by \citet{luh17}, \citet{esp17},
and \citet{luh18} for previously known members and types measured in
this work for new members (Table~\ref{tab:spec}).}
\tablenotetext{d}{* = Discrepant parallax relative to other members of Taurus,
as noted in \citet{luh18} and the Appendix.}
\tablenotetext{e}{1 = 2MASS Point Source Catalog; 2 =  UKIRT Hemisphere Survey;
3 = UKIDSS Data Release 10; 4 = our UKIRT photometry (Sec \ref{sec:ukirtphot});
5 = WIRCam photometry.}
\tablenotetext{f}{nodet = non-detection; sat = saturated; out = outside of the
camera's field of view; bl = photometry may be affected by
blending with a nearby star; bin = includes an
unresolved binary companion; unres = too close to a brighter
star to be detected; false = detection from WISE catalog
appears false or unreliable based on visual inspection.}
\tablenotetext{g}{1 = \citet{luh10}; 2 = \citet{esp14}; 3 = \citet{esp17}; 
4 =  this work.}
\tablenotetext{h}{From \citet{luh10}, \citet{esp14,esp17}, and this work.}
\tablenotetext{i}{$J-H$ and $J-K_s$ = derived from these colors assuming
photospheric near-IR colors \citet{luh10}; CTTS = derived from $J-H$ and $H-K_s$
colors assuming intrinsic colors of classical T Tauri stars from from 
\citet{mey97}; opt spec = derived from an optical spectrum; 
IR spec = derived from an infrared spectrum;
1 = \citet{bri98}; 
2 = \citet{luh00}; 
3 = \citet{str94}; 
4 = \citet{bec07}; 
5 = \citet{whi01}; 
6 = \citet{dew03}; 
7 = \citet{cal04}.
}
\tablecomments{
The table is available in a machine-readable form.}
\end{deluxetable}

\begin{deluxetable}{rrrrcl}
\tabletypesize{\scriptsize}
\tablewidth{0pt}
\tablecaption{Remaining Candidate Members of Taurus from {\it Gaia}\label{tab:cand}}
\tablehead{
\colhead{Gaia DR2 Source Name} &
\colhead{$\alpha$ (J2000)\tablenotemark{a}} & 
\colhead{$\delta$ (J2000)\tablenotemark{a}} & 
\colhead{$G$\tablenotemark{a}} &
\colhead{RUWE\tablenotemark{b}} &
\colhead{Notes}\\
\colhead{} &
\colhead{($\arcdeg$)} &
\colhead{($\arcdeg$)} &
\colhead{} &
\colhead{} &
\colhead{}} 
\startdata
152416436441091584 & 65.288914 & 27.843579 & 16.487 &  2.25 & $0\farcs76$ from M5.25 member {\it Gaia} 152416436443721728 \\
148400225409163776 & 69.936483 & 26.031919 & 20.392 &  1.50 & $3\farcs11$ from M5 member ITG 15 \\
154586322638884992 & 71.227352 & 27.296130 & 12.325 &  1.34 & $1\farcs81$ from K1 member HD 283782 \\
157816855305858176 & 71.458678 & 28.660584 & 6.724 &  2.37 & HD 30111 \\
157816859599833472 & 71.460272 & 28.659692 & 12.123 &  1.17 & $5\farcs97$ from candidate HD 30111 \\
3415706130945884416 & 78.115078 & 22.897051 & 14.303 &  9.74 & $0\farcs64$ from M2.5 member {\it Gaia} 3415706130944329216 
\enddata
\tablenotetext{a}{{\it Gaia} DR2.}
\tablenotetext{b}{\citet{lin18}.}
\end{deluxetable}

\begin{deluxetable}{llll}
\tabletypesize{\scriptsize}
\tablewidth{0pt}
\tablecaption{Observing Log\label{tab:log}}
\tablehead{
\colhead{Telescope/Instrument} &
\colhead{Disperser/Aperture} &
\colhead{Wavelengths/Resolution} &
\colhead{Targets}}
\startdata
HET/LRS2 & VPH grisms/$0\farcs6$ lenslets & 0.65--1.05~\micron/1800 & 2 \\
IRTF/SpeX & prism/$0\farcs8$ slit & 0.8--2.5~\micron/150 & 41 \\
Gemini North/GMOS & R400/$0\farcs5$ slit & 0.6--1.0~\micron/2000 &  21 \\
Gemini North/GNIRS & 31.7 l~mm$^{-1}$/$1\arcsec$ slit & 0.9--2.5~\micron/600 & 51 \\
LAMOST &  540 l~mm$^{-1}$/$3\farcs3$ fiber & 0.37--0.9~\micron/1500 & 31 \\
MMT/Red Channel & 270 l~mm$^{-1}$/$0\farcs75$ slit & 0.58--0.92~\micron/1000 & 34 \\
MMT/Red Channel & 1200 l~mm$^{-1}$/$0\farcs75$ slit & 0.63--0.71~\micron/3250 & 18 \\
MMT/MMIRS & HK grism/$1\farcs2$ slit & 1.25--2.34~\micron/600 & 8
\enddata
\end{deluxetable}

\begin{deluxetable}{lccccc}
\tabletypesize{\scriptsize}
\tablewidth{0pt}
\tablecaption{Spectroscopic Data for Candidate Members of Taurus\label{tab:spec}}
\tablehead{
\colhead{Source Name\tablenotemark{a}} &
\colhead{Spectral} & 
\colhead{$W_{\lambda}$(Li)} &
\colhead{Instrument} &
\colhead{Date} &
\colhead{Member?\tablenotemark{c}} \\
\colhead{} &
\colhead{Type} &
\colhead{(\AA)} &
\colhead{} &
\colhead{} &
\colhead{}}
\startdata
2MASS J04002788+2031591 & M6.5 & \nodata & GNIRS & 2017 Sep 2 & N \\
2MASS J04005482+2117211 & M8 & \nodata & GNIRS & 2017 Sep 27 & N \\
2MASS J04005866+2014043 & M4 & $<$0.2 & SpeX,Red(1200) & 2017 Oct 28,2018 Jan 2 & N \\
UGCS J040132.09+260733.2 & M9.5,M9.5 & \nodata & SpeX,GNIRS & 2017 Oct 30,2017 Sep 29 & Y \\
2MASS J04053214+2733139 & M4.75 & \nodata & Red(270) & 2017 Oct 22 & N?
\enddata
\tablenotetext{a}{Identifications from the 2MASS Point Source Catalog
when available. Otherwise, they based on coordinates from data release 10
of the UKIDSS Galactic Clusters Survey.}
\tablecomments{This table is available in its entirety in a machine-readable
form.}
\end{deluxetable}

\begin{deluxetable}{ll}
\tabletypesize{\scriptsize}
\tablewidth{0pt}
\tablecaption{Disk Fraction for Taurus
\label{tab:disks}}
\tablehead{
\colhead{Spectral Type} &
\colhead{N(II)/N(II+III)}}
\startdata
$<$K6 & 24/33=$0.73^{+0.06}_{-0.09}$ \\
K6--M3.5 & 102/143=$0.71^{+0.03}_{-0.04}$ \\
M3.75--M5.75 & 64/148=$0.43\pm0.04$ \\
M6--M8 & 21/63=$0.33^{+0.06}_{-0.05}$ \\
$>$M8--M9.75 & 15/37=$0.41^{+0.08}_{-0.07}$ 
\enddata
\end{deluxetable}

\clearpage

\end{document}